\documentstyle [preprint,aps,epsf,eqsecnum]{revtex}

\global\arraycolsep=2pt                 

\makeatletter
\def\footnotesize{\@setsize\footnotesize{9.5pt}\xpt\@xpt
\abovedisplayskip 10pt plus2pt minus 5pt
\belowdisplayskip \abovedisplayskip
\abovedisplayshortskip \z@ plus 3pt
\belowdisplayshortskip 6pt plus 2pt minus 2pt
\def\@listi{\topsep 6pt plus 2pt minus 2pt
\parsep 3pt plus 2pt minus 1pt \itemsep \parsep}}
\makeatother
\tightenlines
\def\tab{\hspace{-0.13in}&&\hspace{-0.13in}}

\begin {document}

\preprint {UW/PT-00-03}

\title 
{Finite temperature corrections 
\\
to weak rates prior to nucleosynthesis}

\author 
{Lowell S. Brown}

\address 
{Institute for Theoretical Physics, University of California
\\
Santa Barbara, California 93106
\\
and
\\
Department of Physics, University of Washington
\\
Seattle, Washington 98195}

\author
{R. F. Sawyer}

\address
{Department of Physics, 
University of California
\\
Santa Barbara, California 93106}

\date{\today}

\maketitle

\begin{abstract}

We have reexamined the electromagnetic corrections to the weak
interaction rates for the transformation of neutrons to protons, and
protons to neutrons, in the early universe, before freeze-out. We derive
compact expressions for these rates in terms of thermal expectation
values of products of fields, and we give explicit constructions of
the terms to order $e^2$. We disagree in several respects with results
in the literature.  

\end{abstract}

\newpage

\section{Introduction}

In the early universe, in the era just before nucleosynthesis, the
weak interaction rates combined with the expansion rate determine the
freeze-out neutron-proton ratio at a temperature on the order of 0.3
MeV. This freeze-out ratio in turn essentially determines the
primordial He abundance.\footnote{For a simple pedagogical treatment
of the abundance calculation see Ref. \cite{BBF}.}  In the region of
temperature 0.3 --- 3.0 MeV, in which the weak interaction rates are not
fast enough to preserve chemical equilibrium, it is necessary to
calculate these rates reasonably accurately. The Born approximation
rates, as used by Peebles \cite{peebles} in the first calculations of
the freeze-out ratio and in the later refinements by Wagoner
\cite{wagoner}, nearly suffice for the purposes of this calculation.
But it is possible that the primordial He abundance will be measured
to such a precision that the electromagnetic corrections to the rates
will play a role in the interpretation.

There have been a number of calculations of the electromagnetic
corrections to the freeze-out ratio \cite{dicus} --\cite{esposito}. 
These corrections come from both Coulomb and
radiative effects as modified by the hot plasma in the early universe
and depend on its temperature. Roughly speaking, the method that has
been used in the calculations has been the calculation of all of the T
matrix elements in which $n\rightarrow p$ or $p\rightarrow n$, as
calculated to order $e^2G_W^2$. The squares of these T matrix elements
are then weighted with the appropriate statistical factors for the
incoming and outgoing electrons, positrons, neutrinos, anti-neutrinos
and photons, and integrated over the phase space for the particles. We
call this the ``exclusive channel" approach.

In the present paper we re-address the electromagnetic
correction problem.  Our motivations are several:

1) There is unresolved disagreement in the literature over the correct
treatment of one artifact of the exclusive channel approach, the
so-called ``temperature-dependent wave-function renormalization"
\cite{J} --\cite{sawyer}. And even if the differences with respect to
the the electron's wave function renormalization were resolved, there
is, in exclusive calculations done in Feynman gauge, the necessity of
considering the proton's ``temperature-dependent wave-function
renormalization" as well. This has not been dealt with in the
calculations in the literature, which were all done in Feynman gauge.

2) In the exclusive approach followed by all previous authors, two
reactions have been left out that give contributions of the same order
as the corrections that have been included, namely the reactions
$\nu+e^+ +n \leftrightarrow p+\gamma$. We believe that this is mere
oversight, originating from the fact that these reactions are not
corrections to processes that take place in the absence of the photon
coupling, with a photon added.

3) It would appear that the reported results of the two most complete
reworking of the electromagnetic effects, 
\cite{lopez}, \cite{esposito}, 
do not obey the appropriate detailed balance relation between
proton and neutron rates.

4) The rate correction problem is completely well defined within the
standard model of the weak and electromagnetic interactions, and it is
addressable through perturbation theory. Indeed, for the
temperature-dependent corrections considered in the present paper, the
weak interaction can be taken to be of the local four-Fermi form. We
shall endeavor to present a definitive and final result. In view of
all of the past confusion, we shall take pains to give complete
arguments, to give the analytical results in detail, and to explain
all of the steps.

We shall develop the perturbation theory for the neutron and
proton disappearance rates directly, without going through the
intermediary of the calculation of T matrix elements for exclusive
reactions. Instead, we begin from equations for the inclusive neutron
and proton transformation rates, expressed as the appropriate
integrals over thermal expectation values of electron field
operators. We then apply the standard methods of thermal field theory
to obtain the electromagnetic corrections of order $e^2$ and to reduce
the answers to a form allowing easy computation.  We prove that the
simple detailed balance relation between the proton and neutron rates
remains valid to order $e^2$.  This relation serves as a useful check
on results. There is no role for ``temperature-dependent wave-function
renormalization" in our approach. Our calculation ends with numerical
results that are quite different from those of previous treatments.

We shall assume complete thermal equilibrium for each species. The
effects of a non-thermal neutrino distribution have been dealt
with separately elsewhere\cite{lopez}, \cite{esposito}. 
Their effect on the terms of order $e^2$, a correction
to a correction, is not consequential, particularly so since the 
deviations from a thermal spectrum become appreciable only 
at temperatures very near the freeze-out point.

However the system can be arbitrarily far
from chemical equilibrium. It will suffice to consider a single
neutron in the medium in order to calculate the neutron
transformation rate, and a single proton in the medium to calculate
the proton transformation rate. The electron and neutrino chemical
potentials are taken to vanish. 

\section{General Formulation} 

The rates of neutron and proton appearance and disappearance can be
expressed in terms of the inclusive rates of $\nu_e$ and $\bar \nu_e$
appearance and disappearance. Since we are concerned only with the
modifications of the rates which arise from plasma effects that
involve low energies --- which are `soft processes' --- we may take
the nucleon mass to be infinite\footnote{The zero temperature rates,
which we shall delete from our results, involve a domain of very hard
virtual photons that requires the inclusion of nucleon recoil.  The
zero temperature radiative corrections are summarized in
Ref's. \cite{lopez} and \cite{esposito}.
There are also small temperature-dependent recoil effects
that do not involve the electromagnetic couplings. These are
evaluated in ref. \cite{lopez}.}.  We take the proton or
neutron to be situated at the origin.  We begin by taking the initial
nucleon to be a neutron. As described in detail in Appendix A, the
rate for a neutron to change into a proton may be expressed as
\begin{equation}
\Gamma_n  = { G_W^2 \over (2\pi)^2}  \left( g_V^2 + 3 g_A^2 \right)
\int_{-\infty}^{+\infty} dE_\nu \, E_\nu^2 \,
n(E_\nu) \,  \int_{-\infty}^{+\infty} dt \, e^{+i E_\nu t} \, W(t) \,,
\label{neutron}
\end{equation}
where
\begin{equation}
W(t) =   \left\langle  \psi_e({\bf 0},t) T_-(t) 
\psi_e^\dagger(0) T_+(0) \, \right\rangle_T  \,.
\label{expt}
\end{equation}
Here the angular brackets with the $T$ subscript denote the thermal
average including all plasma and electromagnetic radiative
interactions as well as the single particle state of the (infinitely
heavy) neutron placed at ${\bf r} = 0$. The electron field operator is
denoted by $\psi_e(x)$ and $T_\pm(t)$ are the isospin raising and
lowering operators, with all the operators in the Heisenberg picture
--- their time dependence is controlled by the full Hamiltonian of the
system which also defines the thermal average. We shall assume that
the chemical potentials of all the leptons vanish so that they involve
the Fermi occupation density\footnote{These apply to the free neutrinos
and to the electrons once the electromagnetic corrections appear as a
perturbation.  The electron distribution is, of course, modified by
the electromagnetic interactions, but this is accounted for
automatically in our formalism.} 
\begin{equation}
n(E) = { 1 \over e^{\beta E} + 1 } \,.
\end{equation}
Note that 
\begin{equation}
n(-E) =  e^{\beta E} \, n(E) = 1 - n(E)
\label{block}
\end{equation}
is the Pauli blocking factor for a particle produced with (positive)
energy $E$.

The integral over positive neutrino energies $E_\nu$ in the rate 
$\Gamma_n$ describes neutrinos absorbed from the thermal bath with a
population governed by $n(E_\nu)$. Including corrections of order
$e^2$, it is the sum of the rates for the
processes: 
\begin{enumerate} 
\item{$\qquad \nu+n \rightarrow p+e^{-}$ }
\item{$\qquad \nu+n+\gamma \rightarrow p+e^{-}$ } 
\item{$\qquad \nu+n \rightarrow p+e^{-}+\gamma$ } 
\item{$\qquad \nu+n+e^{+} \rightarrow p+\gamma$}
\end{enumerate} 
The last process in this list, $\nu+n+e^{+}
\rightarrow p+\gamma$, and the reversed process, have been omitted in
all previous calculations in the early universe application. Their
contribution will turn out to be at about the same level as some of
the processes considered in Ref's. \cite{dicus} --\cite{esposito}.

The integral over negative neutrino energies $E_\nu$ in the rate
$\Gamma_n$ describes antineutrinos emitted into the thermal bath with
the Pauli blocking factor $n(- |E_\nu| )$. Again to order $e^2$, it is
the sum of the rates for the processes:
\begin{enumerate}
\item{$\qquad n \rightarrow p+e^{-}+\bar\nu$}
\item{$\qquad n+\gamma \rightarrow p+e^{-}+\bar\nu$}
\item{$\qquad n \rightarrow p+e^{-}+\gamma+\bar\nu$}
\item{$\qquad n+e^{+} \rightarrow p+\bar\nu$}
\item{$\qquad n+e^{+}+\gamma \rightarrow p+\bar\nu$}
\item{$\qquad n+e^{+} \rightarrow p+\gamma+\bar\nu$}
\end{enumerate}

Again as shown in Appendix A, the rate for the disappearance of an 
initial proton is given by
\begin{equation}
\Gamma_p = { G_W^2 \over (2\pi)^2} 
\left( g_V^2 + 3 g_A^2 \right)
\int_{-\infty}^{+\infty} dE_\nu \, E_\nu^2 \,
n(E_\nu) \,  \int_{-\infty}^{+\infty} dt \,  e^{ -i E_\nu t} \, V(t)
\,,
\label{proton}
\end{equation}
where 
\begin{equation}
V(t) =  \left\langle \psi_e^\dagger(0) T_+(0) 
 \psi_e({\bf 0},t) T_-(t) \right\rangle_T  \,.
\end{equation}
The integral over positive $E_\nu$ describes the rates involving the
absorption of antineutrino, the processes in the second list above
with the direction of the reaction arrow reversed. The integral over
negative $E_\nu$ describes the emission of neutrinos, the reverse of
the first list of processes above. For future reference, we note that
the simple change $t \to - t$ gives
\begin{equation}
\Gamma_p = { G_W^2 \over (2\pi)^2} 
\left( g_V^2 + 3 g_A^2 \right)
\int_{-\infty}^{+\infty} dE_\nu \, E_\nu^2 \,
n(E_\nu) \,  \int_{-\infty}^{+\infty} dt \,  e^{ + i E_\nu t} \, 
\tilde V(t)
\,,
\label{pproton}
\end{equation}
where $\tilde V(t) = V(-t)$. 
We make use of the time-translation invariance of the
expectation value to add the time $t$ to all operators to obtain the form  
\begin{equation}
\tilde V(t) =  \left\langle \psi_e^\dagger({\bf 0},t) T_+(t) 
 \psi_e(0) T_-(0) \right\rangle_T  \,.
\end{equation}
This expresses the proton rate $\Gamma_p$ in exactly the same form as
the neutron rate $\Gamma_n$ except for the interchanges
$\psi_e^\dagger \leftrightarrow \psi_e $ and $T_+ \leftrightarrow
T_-$. 

It should be emphasized that our formalism takes account of all these
individual processes and unifies them in terms of a simple
expression. This expression forms the basis in which the plasma
effects on the radiative corrections are easily computed in a complete
and unambiguous fashion. 

To the order $e^2$ to which we work, the neutron and proton proton
rates, $\Gamma_n$ and $\Gamma_p$, obey the detailed balance relation
\begin{equation}
\Gamma_p = e^{ - \beta \Delta } \, \Gamma_n \,,
\label{bal}
\end{equation}
where
\begin{equation}
\Delta = M_n - M_p 
\end{equation}
is the neutron--proton mass difference. Thus only one of the two rates
$\Gamma_n$ or $\Gamma_p$ need be calculated.

To prove the detailed balance statement, we first define the
projection operators $P_n$ and $P_p$ for the point-like heavy neutron
and proton states located at ${\bf r} =  0$.  In terms of these
operators, we have the explicit expressions
\begin{equation}
W(t) =  Z_n^{-1} \, {\rm Tr} \, e^{-\beta H}   \psi_e({\bf 0},t) T_-(t) 
\psi_e^\dagger(0) T_+(0) \, P_n \,,
\end{equation}
and
\begin{equation}
V(t) =  Z_p^{-1} \, {\rm Tr} \, e^{-\beta H} \psi_e^\dagger(0) 
T_+(0)  \psi_e({\bf 0},t) T_-(t) \, P_p \,,
\end{equation}
where
\begin{equation}
Z_n = {\rm Tr} \, e^{-\beta H} P_n \,, \qquad 
Z_p = {\rm Tr} \, e^{-\beta H} P_p \,. 
\end{equation}
The projection operators $P_n$ and $P_p$ are time independent 
(commute with $H$), and they commute with the electron fields, while
\begin{equation}
T_-(t) \, P_p  =  P_n \, T_-(t) \,.
\end{equation}
Using this result, the cyclic symmetry of the trace, and
\begin{equation}
   \psi_e({\bf 0},t) T_-(t) e^{-\beta H} = 
    e^{-\beta H}   \psi_e({\bf 0},t-i\beta) T_-(t-i\beta) \,, 
\end{equation}
we obtain
\begin{eqnarray}
V(t) &=&  Z_p^{-1} \, {\rm Tr} \, e^{-\beta H}   \psi_e({\bf 0}, t-i\beta) 
T_-(t-i\beta) \psi_e^\dagger(0) T_+(0) \, P_n  
\nonumber\\
&=& \left( { Z_n \over Z_p } \right) \, W(t-i\beta) \,.
\end{eqnarray}
Placing this result in the rate formula (\ref{proton}) for the proton
and shifting the time integration variable $t \to t + i\beta $, we
find that
\begin{equation}
\Gamma_p = {Z_n \over Z_p } \, {G_W^2 \over (2\pi)^2} \,
\left( g_V^2 + 3 g_A^2 \right) \,
\int_{-\infty}^{+\infty} dE_\nu \, E_\nu^2 \,
n(E_\nu) \, e^{\beta E_\nu}  \,  
\int_{-\infty}^{+\infty} dt \,  e^{ -i E_\nu t} \, W(t) \,.
\end{equation}
Finally we recall Eq.~(\ref{block}) which states that
 $n(E_\nu) \, \exp\{ \beta \, E_\nu \} = n( - E_\nu)$. Accordingly, we
reflect the neutrino energy integration variable, 
$ E_\nu \to - E_\nu $, to place this result in precisely the form 
(\ref{neutron}) of the neutron rate except for the initial overall 
factor, and hence
\begin{equation}
\Gamma_p = \left( {Z_n \over Z_p} \right) \, \Gamma_n \,.
\end{equation}

In lowest order, 
\begin{equation} 
{Z_n \over Z_p} = e^{ -\beta \Delta } \,, 
\end{equation} 
and the detailed balance statement (\ref{bal}) is obtained to order
$e^2$ if there are no corrections to this ratio to order $e^2$.  It is
a familiar result from statistical mechanics that there is no order
$e^2$ correction to the partition function for a dilute gas. The same
considerations show that there is no $e^2$ correction when a single
charged particle is introduced into a plasma.  This is made explicit
in Ref. \cite{BS} where it is also shown that the first correction is
given by 
\begin{equation} 
{ \delta Z_p \over Z_p } = {1 \over 2} \beta \alpha \, \kappa_D \,, 
\label{debye}
\end{equation} 
where $\kappa_D$ is the Debye wave number. The nature of this result
is clear: The introduction of a charged particle into a plasma alters
its electrostatic field from a pure Coulomb potential $ e / 4 \pi r $
to the Debye screened potential $ (e / 4 \pi r) \exp\{ - \kappa_D \, r
\} $, and thus changes the particle's self-energy by 
\begin{equation}
\delta E = \lim_{r \to 0} {1 \over 2} \, { \alpha \over r }
	\left[ e^{-\kappa_D r} - 1 \right] = - { 1 \over 2 } \,
		\alpha \, \kappa_D \,,
\end{equation}
giving $\delta Z_p / Z_p = - \beta \, \delta E$. For the
electron-positron plasma that is of our concern, the squared Debye
wave number is given by
\begin{equation}
\kappa^2_D = \beta \, e^2  \, \Delta \ell^2 = 
e^2 \left. { \partial \ell \over \partial \mu } \right|_{\mu = 0} \,,
\end{equation}
where $\ell$ is the leptonic charge density,
\begin{equation}
\ell = 2 \int{ (d^3{\bf p}) \over (2\pi)^3 } \left[ 
{1 \over \exp\{\beta [ E(p) - \mu ]\} + 1 } -
{1 \over \exp\{\beta [ E(p) + \mu ]\} + 1 } \right] \,.
\end{equation}
In the high temperature limit in which the temperature is much greater
than the electron mass, $ T \gg m $, this gives $ \kappa_D^2 =
4 \pi \, \alpha \, T^2 / 3 $ and so the plasma correction
(\ref{debye}) is a negligible change 
\begin{equation} 
{ \delta Z_p \over Z_p } = \alpha^{3/2} \, \sqrt{\pi \over 3} 
	\simeq 0.00064 \,.
\end{equation} 
At lower temperatures,
this correction vanishes
exponentially ($ \sim \exp\{ - \beta m \} $) as the temperature
becomes small in comparison with the electron mass, and is thus even
smaller.

To demonstrate the workings of our formalism, we exhibit the rates in
the absence of plasma interactions and radiative corrections.  In this
limit, 
\begin{equation} 
T_-(t) = e^{ +i \Delta \, t } \, T_-(0) \,,
\end{equation} 
and using the free field
correlators (\ref{corr}) for the thermal values of the electron
fields in Eq.~(\ref{neutron}) we get 
\begin{eqnarray}
\Gamma_n^{(0)} &=& G_W^2 \left( g_V^2 + 3 g_A^2 \right) 
{ 1 \over \pi} \int {(d^3{\bf p}) \over (2\pi)^3 } 
\Big[n(-E) \, (E - \Delta)^2 \, n(E-\Delta)
\nonumber\\
&& \qquad\qquad
 +
n(E) \, (E+\Delta)^2 \, n(- E - \Delta ) \Big] \,,
\end{eqnarray}
where we use the short-hand notation $E = E(p)$. 
The first term in the square brackets in the region 
$ m_e < E < \Delta $ describes free neutron decay
$ n \to p + e^- + \bar\nu $ with the final leptons having 
the Pauli blocking factors $n(-E)$, $n(-| \Delta - E |) $. For $ E >
m_e$, the term corresponds to $ \nu + n \to p + e^- $ with the
initial thermal neutrino population governed by $n(E - \Delta)$ and the
final electron having the Pauli blocking factor 
$ n(- E) $. The second term in the square brackets 
represents the process $ e^+ + n \to p + \bar\nu $ with the initial 
positron drawn from the thermal bath with population factor $n(E) $ 
and the final antineutrino having the Pauli blocking factor
$ n(-E - \Delta ) $. In this limit of no corrections, the proton
disappearance rate (\ref{proton}) becomes
\begin{eqnarray} 
\Gamma_p^{(0)} &=& G_W^2 \left( g_V^2 + 3 g_A^2 \right) { 1
\over \pi} \int {(d^3{\bf p}) \over (2\pi)^3 } 
\Big[n(E) \, (E - \Delta)^2 \, n(\Delta-E) 
\nonumber\\
&& \qquad\qquad
+
n(-E) \, (E+\Delta)^2 \, n(E + \Delta ) \Big] \,,
\end{eqnarray} 
with the terms in the square bracket describing the previous neutron
processes in the reverse direction. Since
\begin{equation}
n(\pm E) = e^{ \mp \beta E} \, n( \mp E) \,, \qquad
n(\Delta \mp E) = e^{\pm E} E^{-\beta \Delta } \, n(\pm E - \Delta) 
\,,
\end{equation} 
these rates do indeed obey the detailed balance relation 
$\Gamma_p^{(0)} = \exp\{\beta \Delta \} \Gamma_n^{(0)}$. 
Note that the neutron and proton rates $\Gamma_n$ and $\Gamma_p$ are
related by the interchange\footnote{This equivalence for the
unperturbed rates follows directly from the alternative form
(\ref{pproton}) for $\Gamma_p$. Because of charge-conjugation
invariance, the free-field expectation values of
$\psi^\dagger(t) \psi(0) $ and $\psi(t) \psi^\dagger(0) $ are equal (in
the Majorana representation of the Dirac matrices).  Replacing
$T_-(t)$ by $T_+(t)$ changes the factor $e^{+i \Delta t}$ to 
$e^{- i \Delta t}$. The expectation value of $T_- T_+$ in the neutron
state equals that of $T_+ T_-$ in the proton state.}
$ \Delta \leftrightarrow - \Delta$.

In view of the detailed balance relation (\ref{bal}), we shall
concentrate on the neutron disappearance rate $\Gamma_n$.
To write the lowest-order result in a compact form, we define
\begin{equation} 
\chi(E) = n(E-\Delta) n(-E) (E-\Delta)^2 \,,
\label{chidef}
\end{equation} 
and integrate over the electron solid angle to obtain
\begin{equation}
\Gamma_n^{(0)} =4 \, { G_W^2 \over (2\pi)^3} 
\left( g_V^2 + 3 g_A^2 \right) 
\int_0^\infty  p^2 \, d p \, \left[ \chi(E) + \chi(-E) \right] \,.
\label{freeneutron}
\end{equation} 

The general from for the neutron disappearance rate is  given by
Eq's.~(\ref{neutron}) and (\ref{expt}). Since the rate involves the
operators $ \psi_e({\bf 0},t) T_-(t) $ and 
$ \psi_e^\dagger(0) T_+(0) $ which are electrically neutral (they
commute with the electromagnetic current), the rate is gauge
invariant.  Since we treat the nucleons as being infinitely heavy, it
is convenient to perform calculations in the radiation or Coulomb
gauge, and this we shall do. In this gauge and with a very heavy
proton, the second-order electromagnetic correction involving the 
proton alone is simply a mass shift that is removed by the usual
renormalization.  To the order $e^2$ to which we work, and with the
radiation gauge, the only non-trivial corrections come from the
Coulomb interaction between the electron and proton and the radiative
and Coulomb corrections to the electron (and positron) system.  

We define the $T=0$ rates as the rates that one would get summing the
contributions from all of the processes listed in the introduction,
where thermal distributions are taken for all initial and final
electrons, positrons, neutrinos and antineutrinos, but the temperature
is taken the vanish everywhere else. This completely excludes
participation by the thermal bath of photons. Thus the only real
photon effect in our definition of the $T=0$ rates is a process with a
bremsstrahlung photon in a final state. The $T=0$ terms include all
the radiative and Coulomb corrections that produce ultraviolet
divergences which are removed by the usual renormalization of the
vacuum amplitudes.

We shall subtract all such ``$T=0$'' corrections to define and work
only with temperature-dependent corrections to the rates that are free
of the renormalizations needed to deal with the ultraviolet
infinities.

This definition of $T=0$ corrections agrees with the definition as
used in Ref's. \cite{dicus} --\cite{esposito} except in one regard: If
we consider that part of the reaction rate for the reaction
$n+e^++\nu\rightarrow p + \gamma$ in which the Bose factor for
emission $[1-\exp(-\beta \omega)]^{-1}$ is replaced by unity, this
term falls under the above definition of a T=0 term, but it is {\it
not} included in the $T=0$ terms of
Ref's. \cite{dicus} --\cite{esposito}, for the reason that the process
was not considered at all in these works.

\section{Electron-Proton Interaction}

We shall first look at the plasma correction to the Coulomb
interaction between the electron and proton,
\begin{equation}
H_C(t) = - \int (d^3{\bf r}')(d^3{\bf r}) \, 
\Psi^\dagger_p({\bf r}',t) \Psi_p({\bf r}',t) 
{ e^2 \over 4 \pi  | {\bf r}' - {\bf r} | } 
\psi_e^\dagger({\bf r},t) \psi_e({\bf r},t) \,.
\end{equation}
We start with this piece because it serves as an easy introduction to
the method that will later be applied in the more complex electron
self-energy correction. 
To account for the Coulomb perturbation in first order, we temporarily
continue to imaginary time, $ t \to -i\tau $, with $\tau > 0$.
Then the perturbative rules of thermal field theory in imaginary time 
apply, and the electron-proton Coulomb interaction
correction to Eq.~(\ref{expt}) becomes
\begin{equation}
W_{ep}(-i\tau) = - \int_0^\beta d\tau' 
\left\langle \left( \psi_e({\bf 0},-i\tau) T_-(-i\tau) \, 
H_C(-i\tau') \psi_e^\dagger(0) T_+(0) \, \right)_+ \right\rangle_T  \,,
\end{equation}
where the $( \cdots )_+$ denotes time ordering in the imaginary
time. The operator $H_C(-i\tau')$ makes no contribution when $\tau' >
\tau $ because the ordering places it adjacent to the neutron
state on which it vanishes.  Taking this into account, we may now 
continue back to real time and write
\begin{equation}
W_{ep}(t) = - i \int_0^t dt' 
\left\langle \psi_e({\bf 0},t) T_-(t) \, 
H_C(t') \psi_e^\dagger(0) T_+(0) \,  \right\rangle_T  \,,
\end{equation}
where now no time ordering is involved. Since the perturbation is
explicitly taken into account, we may now use the non-interacting
results. Since the expectation value of 
 $ \Psi^\dagger_p({\bf r}',t) \Psi_p({\bf r}',t) $ is
sharply localized at $ {\bf r}' = 0$, the coordinate ${\bf r}'$ may 
be taken to vanish in the Coulomb potential factor. Hence  
the total charge operator of the proton
appears which acts on a proton state of unit charge, 
and within the expectation value we have
\begin{equation}
T_-(t) \, \int(d^3{\bf r}') 
\Psi^\dagger_p({\bf r}',t) \Psi_p({\bf r}',t) \, \,
 T_+(0) = T_-(t) \, T_+(0) \to e^{+i \Delta \, t } \,.
\end{equation}
Moreover, the electron field thermal expectation value may be 
replaced by the product of free fields defined in Eq.~(\ref{poscorr}) 
of Appendix A. Thus 
\begin{equation}
W_{ep}(t) = + i e^{+i \Delta \, t } \, \int_0^t dt' 
\int(d^3{\bf r}) { e^2 \over 4\pi |{\bf r}| } \,
{\rm tr} \,  S^{(+)}( -{\bf r} , t - t' ) \gamma^0 
	S^{(+)}( {\bf r} , t' ) \gamma^0 \,,
\end{equation}
and using the explicit form (\ref{corr}) for the functions that appear
here,
\begin{eqnarray}
W_{ep}(t) &=& + 4 e^{+i \Delta \, t } \, 
\int{(d^3{\bf p}) \over (2\pi)^3}  {1 \over 2 E}
\int{(d^3{\bf p}') \over (2\pi)^3}  {1 \over 2 E'} \,
{ e^2 \over ( {\bf p} - {\bf p}')^2 } 
\nonumber\\
&& \Bigg\{ { EE' + m^2 + {\bf p} \cdot {\bf p}'  \over E - E' }
\Bigg[ n(-E) n(-E') \, \Big[ \, e^{-i E' t} - e^{- i E t} \, \Big]
\nonumber\\
&& \qquad\qquad 
-  n(E) n(E') \, \Big[ \, e^{i E' t} - e^{ i E t} \, \Big] \Bigg]
\nonumber\\ 
&& \qquad
+ 2 { EE' - m^2 - {\bf p} \cdot {\bf p}'  \over E + E' }
 n(E) n(-E')\, \Big[ \, e^{i E t} - e^{- i E' t} \, \Big] 
	\Bigg\} \,,
\label{funk}
\end{eqnarray}
where $E = E(p)$ and $E' = E(p')$. 

In placing this result in formula (\ref{neutron}) for $\Gamma_n$, the
time integration produce energy-conserving $\delta$ functions. When the
resulting neutrino occupancy factors appear in the form of $n(E -
\Delta)$ multiplying $ n(-E)$ or $n( - E - \Delta)$ multiplying $n(E)$
[or the same forms with $E \to E'$], we have a ``$ T = 0 $''
structure that, as described above, is to removed.  These
are just the neutrino and electron occupancy factors that appear in
the free rate (\ref{freeneutron}).  The terms 
involve a further factor of $n(-E) = 1 - n(E) $ [or $n(-E') = 1 -
n(E')$] with the ``$1$'' parts giving the (Coulomb corrected)
``$T=0$'' part that is to be removed. Our definition thus produces the
temperature-dependent rate correction
\begin{eqnarray}
\Gamma^{(T)}_{n \,, ep} &=& G_W^2 \left( g_V^2 + 3 g_A^2 \right)
{ 2 \over \pi }  
\int{(d^3{\bf p}) \over (2\pi)^3}  {1 \over 2 E}
\int{(d^3{\bf p}') \over (2\pi)^3}  {1 \over 2 E'} \,
{ e^2 \over ( {\bf p} - {\bf p}')^2 } 
\nonumber\\
&& 
\Bigg\{- { EE' +m^2 + {\bf p} \cdot {\bf p}'  \over E - E' }
\Bigg[ n(E) n(-E') \, (E' - \Delta)^2 \, n(E' - \Delta) 
\nonumber\\
&& \qquad\qquad
- n(-E) n(E') \, (E - \Delta)^2 \, n(E-\Delta) 
\nonumber\\
&& \quad
+ \,  n(E) n(E') \left[ (E' + \Delta  )^2 \, n(-E' - \Delta ) -
		(E + \Delta)^2 \, n(-E - \Delta )  \right] \Bigg]
\nonumber\\ 
&& \quad
- 2 \, {EE' -  m^2 -{\bf p} \cdot {\bf p}'  \over E + E' }
 \Big[ n(E) n(E') \, 
  (E + \Delta )^2 \, n(-E -\Delta ) 
 \nonumber\\
&& \qquad\qquad
 + n(-E) n(E') \, (E - \Delta )^2 \, n(E - \Delta ) \Big] \Bigg\} \,.
\label{unsym}
\end{eqnarray}
We make use of the definition (\ref{chidef}) of $\chi(E)$, 
use $pdp = EdE$, and perform the integrals over angles 
to obtain
\begin{eqnarray}
\Gamma^{(T)}_{n \,, ep} &=& -  {e^2 G_W^2 \over (2\pi)^5} 
\left( g_V^2 + 3 g_A^2 \right)
\int_m^\infty dE \,
\int_m^\infty dE' \,
\nonumber\\
&& \qquad
\Bigg\{ {1\over2} \,
 \left[ \left( E + E' \right)^2  
\ln\left( { p + p' \over p - p' } \right)^2 
-4 \, p p' \right]
{1 \over E' - E }
\nonumber\\
&&\qquad\quad
\Big[ n(E') \chi(E) -  n(E) \chi(E') +  n(E') \chi(-E)
	-  n(E) \chi(-E') \Big]  
\nonumber\\
&& \qquad
 + \left[ 4 p p' -  \left( E' - E \right)^2 
\ln\left( { p + p' \over p - p' } \right)^2 \right]
{1 \over E' + E }
\nonumber\\
&& \qquad\qquad\qquad
 n(E') \Big[ \chi(-E) +  \chi(E)  \Big] \Bigg\} \,.
\end{eqnarray}
Each term in the integrand has a Fermi distribution function which 
falls exponentially for large $E$ or $E'$, so that there is no
ultraviolet divergence. The terms in the second square bracket are
odd under the interchange $ E \leftrightarrow E' $ and thus vanish
when $ E = E'$ so as to remove the singularity in the overall
$(E-E')^{-1}$ factor, leaving an integrand that is integrable at 
$E = E'$. Thus we are left with a
well-defined double integral that may be done numerically.  However, 
to save writing, and to place this result in a form that will prove
useful later, we exploit the symmetry of the double integral to
express the result in an asymmetrical form with the understanding that
the potentially singular terms are defined by a principal part
prescription:
\begin{eqnarray}
\Gamma^{(T)}_{n \,, ep} &=& 2 \, {e^2 G_W^2 \over (2\pi)^5} 
\left( g_V^2 + 3 g_A^2 \right)
\int_m^\infty dE \,
\int_m^\infty dE' \, n(E') \,
\Big[ \chi(E) + \chi(-E) \Big]  
\nonumber\\
&& \qquad\qquad
{E \over {E'}^2  - E^2 }
 \left[ 4 p p' - \left(3 {E'}^2 + E^2 \right) 
\ln\left( { p + p' \over p - p' } \right)^2 \right] \,.
\label{elpro}
\end{eqnarray}
It is not difficult to confirm that the principal part prescription is
equivalent to the original symmeterization of the double integral. 

Making use of $n(E) = e^{-\beta E} \, n(-E)$, we can write
the first form (\ref{unsym}) of the result as
\begin{eqnarray}
\Gamma^{(T)}_{n \,, ep} &=& G_W^2 \left( g_V^2 + 3 g_A^2 \right)
{ 2 \over \pi }  
\int{(d^3{\bf p}) \over (2\pi)^3}  {1 \over 2 E} \,
\int{(d^3{\bf p}') \over (2\pi)^3}  {1 \over 2 E'} \,
{ e^2 \over ( {\bf p} - {\bf p}')^2 } 
\nonumber\\
&& 
\Bigg\{- { EE' +m^2 + {\bf p} \cdot {\bf p}'  \over E - E' }
\Bigg[ n(E) \, n(E') 
\nonumber\\
&& \qquad\quad
e^{\beta \Delta} \left\{  (E' - \Delta)^2 \, n(-E' + \Delta) 
- (E - \Delta)^2 \, n(- E + \Delta) \right\}
\nonumber\\
&& \qquad\qquad
+ \left\{ (E' + \Delta )^2 \, n(-E' - \Delta ) -
		(E + \Delta )^2 \, n(-E - \Delta )  \right\} \Bigg]
\nonumber\\ 
&& \quad
- 2 \, {EE' -  m^2 -{\bf p} \cdot {\bf p}'  \over E + E' }
 \Big[n(-E) \, n(E') \Big\{  e^{\beta \Delta} \, 
  (E+ \Delta )^2 \, n(E + \Delta ) 
 \nonumber\\
&& \qquad\qquad\qquad
 +  (E - \Delta )^2 \, n(E - \Delta ) \Big\} \Big] \Bigg\} \,.
\end{eqnarray}
In this guise, it follows immediately that the formal change $\Delta
\to - \Delta $ produces
\begin{equation}
\Gamma^{(T)}_{n \,, ep}( - \Delta) =
e^{-\beta \Delta} \Gamma^{(T)}_{n \,, ep}( + \Delta) \,.
\end{equation}
This is just the detailed balance relation (\ref{bal}) that relates
the neutron and proton rates. Since the subtracted ``$T=0$'' terms are
proportional to the uncorrected rates that obey this relation, we
conclude that the corresponding ``$T \ne 0$'' part of the Coulomb
electron-proton correction to the proton rate may be obtained simply
by the interchange $\Delta \to - \Delta$.  That is: 
\begin{equation}
\Gamma^{(T)}_{p \,, ep} =
\Gamma^{(T)}_{n \,, ep}( \Delta \to - \Delta) 
\label{epsym}
\end{equation}
 
\section{Electron Self-Energy and Radiative Corrections}

We turn now to examine 
the remaining leading-order electromagnetic correction that
involves only the electron (positron) in the neutron rate
defined in Eq's.~(\ref{neutron}), (\ref{expt}).  Since this effect
involves no electromagnetic interaction 
with the proton, we have, effectively, 
\begin{equation}
T_-(t) \, T_+(0) \to e^{i \Delta t} \,.
\end{equation}
It is convenient to 
translate the energy integration variable by $ E_\nu \to E =
E_\nu + \Delta$ to re-express the neutron rate formulas 
(\ref{neutron}), (\ref{expt}) as
\begin{equation}
\Gamma_{n \, {\rm e-e}} = 
{ G_W^2 \over (2\pi)^2 } \left( g_V^2 + 3 g_A^2 \right)
\int_{-\infty}^{+\infty} dE \, (E-\Delta)^2 \, n(E-\Delta) 
\tilde{W}(E) \,,
\label{eerate}
\end{equation}
in which
\begin{equation}
\tilde{W}(E) = \int_{-\infty}^{+\infty} dt \, e^{i E t} 
\,  \left\langle  
\psi({\bf 0},t)   \psi^\dagger(0) \, \right\rangle_T  \,.
\label{realt}
\end{equation}
The angular brackets now denote the thermal average only for 
electrons and photons and, to reduce notational clutter, we have 
removed the $e$ subscript on the electron field operators. 
Note that $E>0$ describes outgoing electrons. 

As we have just seen, the effects of interactions are most
conveniently dealt with in the imaginary time formulation.
However, the continuation to real time is now no longer trivial. 
Hence we pause to review very briefly
the relationship between the unordered real-time function (\ref{realt})
and its counterpart, the time-ordered in imaginary time   
electron thermal Green's function. Since the thermal average 
$ \langle \cdots  \rangle_T$ is the normalized trace
$ Z^{-1} {\rm Tr} \exp\{ -\beta H \} \cdots $, the cyclic symmetry of
the trace may be invoked to interchange the order of operators by
passing then through the evolution operator in imaginary time 
$\exp\{ - \beta H \}$.  Therefore
\begin{equation}
\left\langle \psi({\bf 0},t) \psi^\dagger(0) \, \right\rangle_T =
\left\langle \psi^\dagger(0) \, \psi({\bf 0},t+ i\beta) 
\right\rangle_T \,.
\end{equation} 
In term of Fourier transforms using the sign convention $e^{-iEt}$,
this relates the transforms of the two operator ordering by a factor
of $e^{\beta E}$.  Therefore, 
\begin{equation}
\left\langle \psi({\bf 0},t) \psi^\dagger(0) \, \right\rangle_T =
\int_{-\infty}^{+\infty} { dE \over 2\pi } \, e^{-i E t } \,
 n(-E) \, A(E) \,, 
\label{greater}
\end{equation} 
and
\begin{equation}
\left\langle \psi^\dagger(0) \, \psi({\bf 0},t) \right\rangle_T =
\int_{-\infty}^{+\infty} { dE \over 2\pi } \, e^{-i E t } \,
 n(+E) \, A(E) \,, 
\end{equation} 
where $A(E)$ is the Fourier transform of the thermal average of the
anticommutator and $n(E)$ 
is the Fermi distribution function with vanishing chemical
potential that we have already made much use of. 
The thermal Green's function, time-ordered in imaginary
time, are computed by the usual quantum field theory rules, but in Euclidean 
space, $t \to - i \tau $, and with Bosonic functions periodic and Fermionic
functions anti-periodic in the imaginary time interval $ 0 \,,
\beta$. Thus the electron's thermal Green's function has the Euclidean
4-th momentum component $p_4 = (2m+1) \pi T$, and using the
representation (\ref{greater}), we have
\begin{eqnarray}
\int { (d^3{\bf p}) \over (2\pi)^3}
G({\bf p}, i p_4) \, \gamma^0 &=& \int_0^\beta d\tau \, e^{ip_4 \tau} 
\left\langle \psi({\bf 0},t=-i\tau) \psi^\dagger(0) \, \right\rangle_T 
\nonumber\\
&=& \int_{-\infty}^{+\infty} {dE' \over 2\pi} { A(E') \over E' 
	- i p_4 } \,.
\end{eqnarray}
Continuing back to Minkowski space-time, with 
$ p_4 \to -i E + \epsilon = -i ( E + i \epsilon) $, $\epsilon \to
0^+$, and taking the imaginary part produces
\begin{equation}
 A(E) = 2 \, \int { (d^3{\bf p}) \over (2\pi)^3}
\, {\rm Im} \, G({\bf p}, E + i\epsilon) \, \gamma^0 \,,
\end{equation}
and so, in view of Eq's.~(\ref{realt}) and (\ref{greater}), 
\begin{equation}
\tilde{W}(E) = 2 \, n(-E) \, \int { (d^3{\bf p}) \over (2\pi)^3} 
\, {\rm Im} \, {\rm tr} \, G({\bf p}, E +i \epsilon) \, \gamma^0 \,.
\label{view}
\end{equation}

The thermal electron Green's function has the structure
\begin{equation}
G({\bf p}, ip_4) = \gamma_\mu p_\mu + m + \Sigma(p) \,,
\end{equation}
where $\gamma_4 = -i \gamma^0 $, and we shall compute the self-energy
function $\Sigma(p)$ to first order in $e^2$.  We separate out the
divergent terms that are removed by the renormalization process by
writing 
\begin{equation}
\Sigma(p) = \Sigma^{(0)}(p) + \Sigma^{(T)}(p) \,,
\end{equation}
in which the first term  on the right-hand side is the vacuum function
and the second term vanishes when the temperature vanishes. The vacuum
part gives the renormalization terms.  We avoid dealing with these
renormalization effects by computing only the ``intrinsic''
temperature-dependent corrections, the temperature-dependent
correction for the internal lines of order $e^2$ graphs that have no
divergences.  We continue back to Minkowski space-time and write
\begin{equation}
{ 1 \over \gamma p + m } = { m - \gamma p \over E({\bf p})^2 - 
(E+i\epsilon)^2 } \,,
\end{equation}
where now $\gamma p = \gamma_k \, p_k - \gamma^0 E $, and 
$ E({\bf p}) = \sqrt{ {\bf p}^2 + m^2 } $. To our order,
\begin{eqnarray}
{\rm Im} \, G({\bf p}, E + i\epsilon) &=& - ( m - \gamma p ) 
\Bigg\{ { 1 \over \left[ E({\bf p})^2 - E^2 \right]^2 } 
\, {\rm Im} \, \Sigma({\bf p}, E + i\epsilon) 
\nonumber\\
&+& 
\Sigma^{(T)}({\bf p},  E ) 
\, {\rm Im} \, { 1 \over \left[ E({\bf p})^2 - (E+i\epsilon)^2 \right]^2 } 
\Bigg\} (m - \gamma p ) \,.
\label{part}
\end{eqnarray}
In the second term on the right here, only the temperature-dependent
part of the self-energy function enters --- the vacuum contribution is
removed by renormalization. To deal with this second  term, we note
that
\begin{eqnarray}
{\rm Im} \, { 1 \over \left[ E({\bf p})^2 - (E+i\epsilon)^2 \right]^2 } 
&=& {1 \over 2 E } { \partial \over \partial E }
\, {\rm Im} \, { 1 \over E({\bf p})^2- (E+i\epsilon)^2  }  
\nonumber\\  
&=&
{1 \over 2 E } { \partial \over \partial E } \, {\pi \over 2 E({\bf p}) } \,
\left[ \delta \left( E - E({\bf p}) \right) -  
\delta \left( E + E({\bf p}) \right) \right] \,.
\label{derive}
\end{eqnarray}
 
To summarize, we recall the definition (\ref{chidef}) of $\chi(E)$, 
and use Eq's.~(\ref{derive}), (\ref{part}), and (\ref{view}) in the
rate formula (\ref{eerate}) to obtain
\begin{eqnarray}
\Gamma_{n \, {\rm e-e}} &=& 
{ G_W^2 \over (2\pi)^2 } \left( g_V^2 + 3 g_A^2 \right)
\int_{-\infty}^{+\infty} dE \, \int { (d^3{\bf p}) \over (2\pi)^3} \,
{\rm tr} \, \gamma^0 \, \Bigg\{
\nonumber\\
&& \quad
 - ( m - \gamma p ) 
 { \chi(E) \over \left[ E({\bf p})^2 - E^2 \right]^2 } \, 
2 \, {\rm Im} \, \Sigma({\bf p},  E + i\epsilon) ( m - \gamma p ) 
\nonumber\\
&& \quad
+ \left[ \delta \left( E - E({\bf p}) \right) -  
\delta \left( E + E({\bf p}) \right) \right]
\nonumber\\
&& \quad
\times { \pi \over 2 E({\bf p} ) } \, { \partial \over \partial E } \, 
\left[  E^{-1} \chi(E) \,  (m - \gamma p ) 
\Sigma^{(T)}({\bf p}, E ) (m - \gamma p ) \right] \Bigg\} \,.
\label{wonder}
\end{eqnarray}
In the last line, the derivative of the delta function has been
integrated by parts. When this derivative acts upon $ \chi(E)$, the
remaining factor is gauge invariant, and the result corresponds to a
temperature-dependent shift of the electron's energy in the plasma. The
terms that result when the derivative acts on the remaining factors
could be associated with some sort of ``temperature-dependent
wave-function renormalization'', but these terms are not gauge
invariant, and so they cannot be given any independent physical
interpretation. The first line, involving the imaginary part of the 
complete self-energy function describes bremsstrahlung processes in the
plasma and, since an imaginary part is taken, it has no ultraviolet
divergences. The first line does, however, contain ``$T=0$''
contributions that must be removed.  This is discussed in the next
section and in more detail in Appendix B.

It should be {\em emphasized} that the electron-electron contribution
(\ref{wonder}) given here, together with the electron-proton contribution
(\ref{elpro}) given at the end of the previous section, give a
complete and unified description of {\em all} the thermal, plasma
effects on the leading electromagnetic corrections to the neutron
disappearance rate.  Some distinct and separate physical processes
may be identified in the total rate contribution such as the energy
shift mentioned above which will soon be described in more detail.
However, other terms, such as those that may be associated with a `wave
function renormalization' contribution, are not gauge invariant and
have no physical meaning whatsoever. Indeed, the electron-proton 
Coulomb correction computed in the previous section is not gauge
invariant and has no physical significance by itself.  We turn in the
next section to assemble our results in terms of physically
significant parts.  We do this so as to obtain formulae that are
easily read and not
so long as to tire the eye, but still have well-defined physical
significance. The details of the calculation of these
results appears in Appendix B.

Before going into the details of our results in the next section, we
note that the alternative form (\ref{pproton}) shows that the proton rate 
may be obtained from the neutron rate by the interchanges $\psi 
\leftrightarrow \psi^\dagger$ and $ T_+ \leftrightarrow T_-$. Since 
$\Gamma_{\rm e-e}$ involves no interaction with the proton, charge
conjugation invariance holds for the electron field thermal
expectation values, and so they are not altered by the interchange 
$ \psi \leftrightarrow \psi^\dagger $. The interchange 
$ T_+ \leftrightarrow T_-$ just changes the associated time dependence
from $e^{i \Delta t }$ to $e^{ - i \Delta t}$. Hence the proton rate is
related to the neutron rate by
\begin{equation}
\Gamma_{p \, {\rm e-e}} = \Gamma_{n \, {\rm e-e}}(\Delta \to - \Delta)
\,.
\label{eesym}
\end{equation}

\section{Results}

As we have just noted, the first line of Eq.~(\ref{wonder}) describes
real photon processes.  One piece involves a factor of the thermal
photon distribution function
\begin{equation}
f(k) = { 1 \over e^{\beta k} -1 } \,,
\end{equation}
and the remaining part does not involve this function. This remaining
part has pieces corresponding to the real photon emission processes
$ \nu + n \to p + e^- + \gamma $, $ \nu + e^+ + n \to p  + \gamma $,
$ n \to p + \bar\nu + e^- + \gamma $, and
$ e^+ + n \to p + \bar\nu + \gamma $. As described in Appendix B,
these pieces correspond to parts of the ``$T=0$'' rate that are easily
identified and removed to produce the desired $T \ne 0 $ rate
contribution. 
The remaining part of the real photon processes described by the
first line of Eq.~(\ref{wonder})  is the
photon thermal bath contribution  (\ref{yetinfr}) computed in Appendix
B. This result reads 
\begin{eqnarray}
\Gamma_{n \, {\rm e-e}}^{(\gamma,\gamma)} &=& 
2\, {e^2  G_W^2 \over (2\pi)^5 } \left( g_V^2 + 3 g_A^2 \right)
 \int_0^\infty {dk \over k}  \, f(k) \, \int_m^\infty d{\cal E}  
\nonumber\\
&& 
\Big[
\left[ n(-{\cal E}) \tilde\chi({\cal E} - k) + 
 n({\cal E}) \tilde \chi(- {\cal E} + k) \right] F_-
+ \left[ n(-{\cal E}) \tilde\chi({\cal E} + k) +  
	 n({\cal E}) \tilde\chi(- {\cal E} - k) \right] F_+
	\Big] \,.
\nonumber\\
&&
\label{realr}
\end{eqnarray}
Here
\begin{equation}
\tilde \chi(E) = n(E-\Delta) \, (E-\Delta)^2 \,,
\end{equation}
according to Eq.~(\ref{tildechi}). The functions
\begin{equation}
F_\pm = A \pm k \, B 
\end{equation}
appear in Eq.~(\ref{fpm}), with
\begin{equation}
A = 
\left[ 2 \, {\cal E}^2 + k^2 \right]
	\ln\left({ {\cal E} + q \over {\cal E} - q } \right)
		- 4 \, q \,  {\cal E}  \,,
\end{equation}
and
\begin{equation}
B =  2 \, {\cal E} \, 
	\ln\left({ {\cal E} + q \over {\cal E} - q } \right)
		- 4 \, q  \,,
\end{equation}
where
\begin{equation}
q = \sqrt{ {\cal E}^2 - m^2 } \,.
\end{equation}

Although this set of terms in Eq.~(\ref{realr}) has no ultraviolet 
divergences, since $f(k) \, k^{-1} \to
k^{-2}$ as $k \to 0$, it is infrared divergent.  This contribution is
rendered finite in the infrared by a part of the contribution
contained in the last two lines of
Eq.~(\ref{wonder}) \footnote{ The fact that the infrared divergences must cancel, order by order, for these rate calculations was proved in \cite{baier}.}. It is one of the parts that arises from
all the terms that remain when the energy derivative $\partial /
\partial E$ does not act on $\chi(E)$. This piece, which  
involves an overall factor of the photon distribution function $f(k)$,
is calculated in Eq.~(\ref{photon}) of Appendix B to be
\begin{equation}
\Gamma_{n \, {\rm e-e}}^{\gamma,Z} = 
2\, {e^2  G_W^2 \over (2\pi)^5 } \left( g_V^2 + 3 g_A^2 \right)
 \int_0^\infty {dk \over k} \, f(k) \, \int_m^\infty d{\cal E} 
\, A \, \left\{ - 2  n(-{\cal E}) \, \tilde\chi({\cal E}) 
	- 2  n({\cal E}) \, \tilde\chi(- {\cal E}) \right\} \,.
\label{photonn}
\end{equation}
Adding Eq.~(\ref{photonn}) to the real photon rate (\ref{realr}),
and writing our rationalized charge in terms of the fine structure
constant, $\alpha = e^2 / 4\pi$, gives
\begin{eqnarray}
\Gamma_{n \, {\rm e-e}}^{(\gamma,f)} &=& 
{2 \, \alpha \over \pi } { G_W^2 \over (2\pi)^3 } 
\left( g_V^2 + 3 g_A^2 \right)
 \int_0^\infty {dk \over k}  \, f(k) \, \int_m^\infty d{\cal E}  
\Bigg\{
\nonumber\\
&& \qquad
A \, 
\Bigg[  n(-{\cal E}) \left\{\tilde\chi({\cal E} - k) +  
\tilde\chi({\cal E} + k)  - 2 \tilde\chi({\cal E} ) \right\}
\nonumber\\
&& \qquad\qquad\qquad\qquad
+  n({\cal E}) \left\{ \tilde\chi(-{\cal E} + k) + 
	\tilde\chi(- {\cal E} - k)
 - 2  \tilde\chi(-{\cal E}) \right\}  \Bigg]
\nonumber\\
&& \quad
 - k B \, 
\Bigg[n(-{\cal E}) \left\{ \tilde\chi({\cal E} - k) 
-  \tilde\chi({\cal E} + k) \right\}
 +n({\cal E}) \left\{ \tilde\chi(-{\cal E} + k) -
  \tilde\chi(- {\cal E} - k) \right\} \Bigg]
\Bigg\} \,.
\nonumber\\
&&
\label{infrafine}
\end{eqnarray}
The weight $k^{-1} f(k)$ diverges as $k^{-2}$ when $k$ vanishes, 
$ k \to 0 $.  The set of terms in the first square brackets is of
order $k^2$ for small $k$. The terms in the final square brackets are
of order $k$ for small $k$, but they are multiplied by an explicit
factor of $k$.  Hence, this contribution to the neutron rate has no
infrared divergence.  The $k$ and ${\cal E}$ integrals are damped for
large $k$ and ${\cal E}$ by the Bose and Fermi phase-space density
functions, and so it is also well behaved in the ultraviolet as well
as the infrared.  Since this is the only part of the rate that is
proportional to the photon density $f(k)$, it is a well-defined, 
gauge invariant part of the neutron disappearance rate.

The final pieces that arise when the energy derivative $\partial /
\partial E$ does not act on $\chi(E)$ in the 
last two lines in Eq.~(\ref{wonder}) is presented in Eq.~(\ref{waveg})
in Appendix B.  Adding this to the modification (\ref{elpro}) of the 
neutron rate brought about by the Coulomb interaction between the
electron and proton and again using $ \alpha = e^2 /4\pi$ gives
\begin{eqnarray}
\Gamma_{n \,,\, ep+ee} 
&=&
{2 \, \alpha \over \pi} { G_W^2 \over (2\pi)^3 } \, 
\left( g_V^2 + 3 g_A^2 \right)
\int_m^\infty dE \, \Big[ \chi(E) + \chi(-E) \Big] \,
\int_m^\infty dE' \,  n(E') \, 
\nonumber\\
&& \qquad 
 { E \over {E'}^2 - E^2 }
\Bigg\{ 8\,  p p' - \left[ {E'}^2 +  E^2 \right]
  \, \ln\left( { p+p' \over p-p' } \right)^2 
\nonumber\\
&& \qquad\qquad
-  {E' \over E }  \,  \left [ {E'}^2 + E^2 \right]
 \, \ln \left( { \left[ E E' + pp' \right]^2 - m^4 \over
	\left[ E E' - pp' \right]^2 - m^4 } \right) 
\Bigg\} \,.
\label{invar}
\end{eqnarray}
Since the only remaining term involves the gauge-invariant energy
shift shown in Eq.~(\ref{shiftt}) below, and all the other previous 
contributions are gauge invariant, we
conclude that Eq.~(\ref{invar}) is a well-defined, gauge-invariant
correction to the neutron disappearance rate.  
Just as we have done in the writing of Eq.~(\ref{elpro}), it is left
implicit that the potentially singular terms are defined by a principal
part prescription.  

Finally, there is the term when the energy derivative in the last line
of Eq.~(\ref{wonder}) acts on $\chi(E)$. In this case, the electron
self-energy function is evaluated on mass shell and thus describes a
gauge-invariant, temperature-dependent energy shift correction which,
after performing the angular integrations, has the form
\begin{eqnarray}
\Gamma_{n \, {\rm e-e}}^{(\Delta E)} &=& 
4 \, {G_W^2 \over (2\pi)^3}  \left( g_V^2 + 3 g_A^2 \right)  
\int_{-\infty}^{+\infty} dE \, \int_0^\infty p^2 \, dp
\nonumber\\
&& \qquad
\left[ \delta \left( E - E({\bf p}) \right) +  
\delta \left( E + E({\bf p}) \right) \right] \,
\Delta E^{(T)}({\bf p}) \,
{ \partial \over \partial E } \,  \chi(E)  \,.
\label{shiftt}
\end{eqnarray}
As explained in more detail in Appendix B, the energy shift 
$\Delta E^{(T)}({\bf p})$ is a shift in the position of the pole in
the interacting thermal electron propagator --- it gives the energy
shift to create an electron (positron) of momentum ${\bf p}$ in the 
plasma relative to the energy needed to create the particle in the
vacuum. The form of Eq.~(\ref{shiftt}) is precisely the change in the
free rate (\ref{freeneutron}) when the energy--momentum relation of
the electron (positron) is altered. The energy shifts is computed in
Appendix B. The results displayed in
Eq's.~(\ref{gamshift}) and (\ref{elshift}) give
\begin{eqnarray}
\Delta E^{(T)}({\bf p})  &=& 
{2 \alpha \over \pi  E } \left\{ \int_0^\infty k^2 \, dk 
 \, { 1 \over k} \, f(k) 
+ \int_0^\infty q^2 \, dq
 \, { 1 \over E({\bf q}) } \, n(E({\bf q})) \, 
\Bigg[ 1 - { m^2 \over 4 \, q  p }
\ln \left( { q + p \over q - p } \right)^2 \Bigg] \right\} \,. 
\nonumber\\
&&
\label{eeesh}
\end{eqnarray}
The energy
derivative in the energy-shift term (\ref{shiftt}) may be
removed by an integration by parts using $ \partial / \partial E 
= (E/p) \, \partial / \partial p $. Using the explicit form
(\ref{eeesh}) of the energy shift, we find that 
\begin{eqnarray}
\Gamma_{n \, {\rm e-e}}^{(\Delta E)} 
&=&
- { 8 \, \alpha \over \pi }  \, { G_W^2 \over (2\pi)^3 } \, 
\left( g_V^2 + 3 g_A^2 \right)
\int_m^\infty { E \over p} \, dE \,  
\Big[ \chi(E) + \chi(-E) \Big] 
\nonumber\\
&& \qquad  \left\{
\int_0^\infty dk \, k \, f(k) + \int_m^\infty  dE' \, p' \, n(E')
\left[ 1 -  
 { m^2 \over {E'}^2 - E^2 } \right] \right\} \,.
\label{done}
\end{eqnarray}
Again it has been left implicit that integrand in the second double 
integral in $E$ and $E'$ is defined by the principal part prescription. 

The non-trivial plasma effects on the one-loop electromagnetic
correction to the total neutron disappearance rate is the sum of the
results that we have enumerated:
\begin{equation}
\Delta \Gamma^{(T)}_n = 
	 \Gamma_{n \, {\rm e-e}}^{(\gamma,f)} 
	+ \Gamma_{n \,,\, ep+ee} 
	+ \Gamma_{n \, {\rm e-e}}^{(\Delta E)} \,,
\label{total}
\end{equation}
in which the successive contributions are given in 
Eq's.~(\ref{infrafine}), (\ref{invar}), and 
(\ref{done}). We should emphasize one last time that this result
follows from a straight forward evaluation of the corrections to the
basic rate formulas (\ref{neutron}) and (\ref{expt}). We have broken
up the complete result into smaller pieces in large part so as to have
reasonably short expressions that do have gauge-invariant, physical
meanings. But we have not patched together a set of various physical
processes. Rather, we have started from a well-defined general
formulation and just worked out the consequences.

The second set of terms on the right-hand side of Eq.~(\ref{total}) 
involve singular integrands that require principal part definitions. 
To write the result in a less singular form, we 
perform some algebraic rearrangement%
\footnote{An ingredient needed for this is
the identity
$$
{ \left[ E E' + pp' \right]^2 - m^4 \over
	\left[ E E' - pp' \right]^2 - m^4 }  =
\left( { E E' + pp' + m^2 \over
	 E E' - pp' + m^2 } \right)^2 \,
\left( { p + p' \over p - p' } \right)^2  \,.
$$\vspace*{-18pt}} 
and partial integrations to express the sum of these two terms in the
form
\begin{eqnarray}
\Gamma_{n \,,\, ep+ee} + \Gamma_{n \, {\rm e-e}}^{(\Delta E)} 
&=& {2 \, \alpha \over \pi} { G_W^2 \over (2\pi)^3 } \, 
\left( g_V^2 + 3 g_A^2 \right)
\int_m^\infty dE \, \Big[ \chi(E) + \chi(-E) \Big] \,
\nonumber\\
&& \qquad\qquad \left\{
- { 2 \pi^2 \over 3 \beta^2} \, { E \over p} + \int_m^\infty dE' \,
F(E,E') \right\} \,,
\label{nosing}
\end{eqnarray}
with
\begin{eqnarray}
&& F(E,E') = - { 1 \over 4} 
\ln^2 \left( { p+p' \over p-p' } \right)^2 \Bigg\{
n'(E') \, { p' \over p} \, {E^2 \over E'} \left[ E + E' \right]
+ n(E') \, { E^2 \over p p'} \, 
\left[ E' + {m^2 E \over {E'}^2 } \right]
\Bigg\}
\nonumber\\
&& \qquad\qquad\quad  
+ \ln \left( { p+p' \over p-p' } \right)^2 \Bigg\{
n'(E') \left[ {p'}^2 {E \over E'} \left( { m^2 \over p^2} + 2 \right) 
 -  E^2 \, {p' \over p } \, L(E,E') \right]
\nonumber\\
&& \qquad\qquad\qquad\quad + \, n(E') 
\left[ {  E m^2 \over p^2 {E'}^2 } \left(
{E'}^2 {+} 2p^2 {+} m^2 \right) - { E^2 + {E'}^2 \over E+E' } 
- { E^2 E' \over p p' }  L(E,E') \right] \Bigg\}
\nonumber\\
&& \qquad\qquad\quad
- \, n(E') \Bigg\{ 4 E \, { p' \over p } + 2 E' \, L(E,E')  \Bigg\} \,.
\end{eqnarray}
Here in the first equality we have explicitly evaluated the simple
integration over the photon distribution $f(k)$ in Eq.~(\ref{done}),
and in the second equality we have used the definitions
\begin{equation}
n'(E') = {d n(E') \over d E' } \,,
\end{equation}
and
\begin{equation}
L(E,E') =    \ln \left( { E E' + pp' + m^2 \over
	 E E' - pp' + m^2 } \right) \,.
\end{equation}
This non-singular expression of these rates is a convenient form to
use in numerical evaluations as we shall do below.   

It was shown in Eq.~(\ref{epsym}) that the changes in the neutron and
proton rates brought about by the electron-proton Coulomb interaction
are related by the formal substitution $\Delta \leftrightarrow -
\Delta$. The same substitution relates the total electron self-energy
and radiative corrections to rates as shown in Eq.~(\ref{eesym}).
A glance at the details of the ``$T=0$'' subtractions made to the
electron self-energy and radiative corrections shows that those for
the proton and neutron are also related by the interchange 
$\Delta \leftrightarrow - \Delta$. We therefore conclude that the
complete ``$T \ne 0$'' corrections to the proton rate are given by
\begin{equation}
\Delta \Gamma^{(T)}_p = 
\Delta \Gamma^{(T)}_n(\Delta \to - \Delta) \,.
\end{equation}

As was discussed in the Introduction, the process 
$\nu + e^+ + n \to p + \gamma $ has been hitherto inadvertently
omitted in the literature. This contribution to the neutron rate is
identified in Eq.~(\ref{forgive}), which we repeat here:
\begin{eqnarray}
\Gamma_{n \, {\rm F}} &=& 
{2 \, \alpha \over \pi}  {G_W^2 \over (2\pi)^3} 
\left( g_V^2 + 3 g_A^2 \right)
\, \int_m^\infty d{\cal E}  \,  n({\cal E}) 
 \int_{{\cal E} + \Delta}^\infty {dk \over k} \, 
\left[ -f(-k) \right] \,
\tilde\chi(-{\cal E} + k ) \, F_- \,.
\label{fforgive}
\end{eqnarray}
As discussed further in Appendix B, the $T=0$ part of this 
rate is obtained by taking $-f(-k) \to 1$,
\begin{eqnarray}
\Gamma_{n \, {\rm F}}^{(T=0)} &=& 
{2 \, \alpha \over \pi}  { G_W^2 \over (2\pi)^3} 
\left( g_V^2 + 3 g_A^2 \right)
\, \int_m^\infty d{\cal E}  \,  n({\cal E}) 
 \int_{{\cal E} + \Delta}^\infty {dk \over k} \, 
\tilde\chi(-{\cal E} + k ) \, F_- \,.
\label{repeatforgot}
\end{eqnarray}

\section{Discussion}

We have derived expressions for the rate shifts, to order $e^2$, in
terms of thermal expectation values involving only electron field
operators. The perturbation expansion of these expectation values has
been carried out using the standard methods of thermal field
theory. The ``$T=0$" terms have been carefully defined and removed.
The remainder has been expressed in the form of convergent
two-dimensional integrals.

We gave formulae for only the neutron rates $\Gamma_n$; as we stressed
earlier, if we had calculated the whole rate, rather than the finite
temperature piece, we could use detailed balance directly to get a
proton rate $\Gamma_p$ from $\Gamma_n$.  But since the present work is
limited to the finite temperature part, we use the transformation
$\Delta\rightarrow -\Delta$ in the integrands to get the corrections
to the proton rate $\Delta \Gamma_p^{(T)}$ from those for the neutron
rate.  We have performed the integrals in Eq's.~(\ref{infrafine}),
(\ref{invar}), and (\ref{done}) to compute the total
temperature-dependent part of the leading electromagnetic correction
to the neutron rate as shown in Eq.~(\ref{total}). Their counterparts
for the proton process have been computed in the same way after making
the substitution $\Delta \to - \Delta$.  We exhibit the results of the
integration in the form of the fractional changes of the neutron rate
$\Delta \Gamma_n^{(T)} / \Gamma_n^{(0)}$ and proton rate $\Delta
\Gamma_p^{(T)} / \Gamma_p^{(0)}$, where $\Gamma_n^{(0)}$ is the free
neutron rate given in Eq.~(\ref{freeneutron}), while the free proton
rate $\Gamma_p^{(0)}$ is obtained by the same calculation after the
making the replacement $\Delta \to - \Delta$. The results are shown as
a function of temperature in Fig.~1.

We cannot compare these results directly with those in the literature,
since previous authors have omitted the processes $n+e^+ +
\nu\rightarrow p+\gamma $ and $p+\gamma\rightarrow n+e^+ + \nu $.  The
first of these processes, since it has an outgoing photon, generates a
$T=0$ part that has been subtracted as a part of the general $T=0$
subtractions in (\ref{total}). We define a ``revised" $T$ part of the
rate as that which must be added to the previous literature's $T=0$
part to get the complete answer. The ``revised" $T$ neutron rate is
found by adding the contribution (\ref{repeatforgot}) to the the $T$
rate plotted in Fig.~1. There is no $T=0$ piece of the reaction,
$p+\gamma\rightarrow n+e^+ + \nu $, since here the photon is in the
initial state, and therefore the proton``revised" T rate is the same
as that plotted in Fig.~1.

We plot the fractional ``revised" rate corrections in Fig.~2. For
comparison we also show data reconstructed from the figures of
\cite{lopez}, shown as dashed curves; the latter are derived by
combining the numbers shown in Fig.~11 of \cite{lopez}, called
the``finite T corrections", with those of Fig.~16 of that reference,
called the ``finite temperature electron mass corrections."  We see
that at a temperature of 0.3 MeV our results nearly coincide with those
of \cite{lopez}, but that the two families of results diverge rapidly
at higher temperatures. We find positive rate corrections that are
larger than those of Ref. \cite{lopez} for both processes, and we find
much less difference between the neutron and proton corrections. Both
of these changes lead to faster equilibration.

Since the electromagnetic corrections are quite small, the detailed
balance relation (\ref{bal}), expressed in terms of fractional
corrections, is simply 
\begin{equation} 
(\Delta \Gamma_p^{(T)} +\Delta
\Gamma _p^{(T=0)}) / \Gamma_p^{(0)} = 
(\Delta \Gamma_n^{(T)} +\Delta \Gamma_n^{(T=0)} )/ \Gamma_n^{(0)} \,.
\label{rebalance} 
\end{equation} 
We can, in principle, try to check our answers with this statement by
adding to our results the $T=0$ results of \cite{lopez}, shown in
Fig.~8 of this reference, to our finite $T$ corrections.
Unfortunately this gets us into a spot of guesswork, since we cannot
read the $\Delta \Gamma _p^{(T=0)} - \Delta \Gamma _n^{(T=0)}$ 
differences
off of this plot quite precisely enough to get a clear answer. It
would appear that these differences, as shown in the plot, are a
little too small to compensate for the 
$\Delta \Gamma _p^{(T)}-\Delta\Gamma _n^{(T)}$ 
differences that we calculate and that are plotted in
our Fig.~1. The same detailed balance test using the finite $T$
results of \cite{lopez}, which shows a larger difference in the $n$ and
$p$ corrections, appears to fail by more. We conclude that the
approximations that were made in the calculation of the $T=0$ terms in
Ref's. \cite{dicus} and \cite{lopez} may deserve some further scrutiny.

It is harder to make a detailed comparison of our results with those
of Ref. \cite{esposito} because of the way the latter are
presented. However in Fig.~12 of Ref. \cite{esposito} we see large
violations of the detailed balance requirement (\ref{rebalance}).

We believe that with our formalism we have avoided the questions that
have led to confusion and ambiguity in the literature. We should
comment a little further on the wave-function renormalization
questions that have plagued previous calculations. In the T matrix
approach, with exclusive channels, the need for this renormalization
arises from processes on an external line. Conceptually at least, one
needs to sum the bubbles on the external line and calculate therefrom
the residue of the energy shifted pole.  The renormalization constant
so obtained can then be expanded to second order in $e$. The
complication in the case of finite temperature, with its preferred
coordinate system, is that the multiplicative factor contains a
transformation on the Dirac spinor indices, as worked out in
\cite{sawyer}. It is instructive to consider how these considerations
would be embodied if we separated our result (\ref{part}) into
exclusive channels. The ``external line" problem would present itself
if we tried to evaluate the imaginary part of the square of the
electron propagator, encountered in the third line of Eq.~(\ref{part})
(with the omission of the spinor numerators) as ${\rm Im}[E( {\bf
p})^2-(E+i \epsilon)^2]^{-2}$, by taking the imaginary part of one
factor (a delta function) times the real part of the other (a
denominator that is singular at the point of vanishing of the delta
function).  This is the structure that comes from taking a Feynman
graph with a correction on an external line. We can see some of the
benefits of the inclusive approach from the operations that follow our
Eq.~(\ref{part}), in which the representation of the imaginary part given
by Eq.~(\ref{derive}) avoids entirely the $i\epsilon$ delicacies required
in treating exclusive channels.

But there appears to be a deeper problem with the methods that have
been used in the literature. In our Eq.~(\ref{part}) the propagators in
question are the ordinary real time vacuum functions; the thermal
effects have all been incorporated through the preceding
formalism. However, in \cite{dicus} and \cite{cambier} as well as in
the subsequent papers on this subject, the individual graphs were
calculated with real time thermal propagators of the form (for
electrons)
\begin{equation} 
S(p) = (\gamma p + m)^{-1} + 2\pi
n(|E|) (m - \gamma p ) \delta(p^2+m^2) \,. 
\end{equation}
When this propagator gets squared, it is truly meaningless; the
interpretation is no longer just a question of following the $i
\epsilon$'s carefully. It appears to us that the square of the delta
function has been implicitly treated in an arbitrary way in the
treatments based on real time thermal propagators.  This defect in the
the real-time approach has been noted before in other contexts, for
example in \cite{kapusta} and \cite{dolan}.

Finally, as we remarked earlier, the wave function renormalization is
gauge dependent. In Feynman gauge, in which photons are emitted and
absorbed by the proton, there is a temperature dependent part of the
proton wave-function renormalization that must be calculated in order
to get the correct answer in the exclusive approach. This appears to
be missing in all of the published calculations. We conclude, from the
above discussion, that it would be futile to try to make term by term
comparisons of our results with the previous ones.
However, we can say that most of the discrepancy comes from
sources other than the new channel that we noted, $\gamma+p
\leftrightarrow  \nu +e^++n$ . This fact is made
clear by Fig.~3, which shows the complete $T=0$ + $ T \ne 0$
contributions from these reactions. Since these terms obey
detailed balance by themselves, there is no difference between
the neutron and the proton fractional corrections shown in Fig.~3.

A further use for the formalism developed in this paper would be in
the calculation of the electromagnetic corrections to the charged
current neutrino opacity in the environment of the supernova core.
The neutrino reactions to be considered are the same ones that we have
dealt with in the present paper.  But the electrons are very
degenerate, with the electron chemical potential in the 10's of
MeV. Hardy \cite{hardy} has found, using some of the technology from
the early universe calculations that we have discussed above, some
evidence of significant electromagnetic effects in the rates. To get
the complete answer to order $e^2$ one can adapt our equations to get
the neutrino absorption rate in the presence of a non-vanishing
electron chemical potential.

\section* {Acknowledgments}

The work of L. S. B. was supported, in part, by the U.S. Department of
Energy under Grant No.  DE-FG03-96ER40956; that of R. F. S. was
supported, in part, by the National Science Foundation under the grant
PHY-9900544.  One of the authors (R. F. S.) wishes to thank the
Institute for Nuclear Theory at the University of
Washington, where some of this work was carried out.

\newpage

\begin {figure}[ht]
    \begin{center}
        \epsfxsize 5.6in
        \begin{tabular}{rc}
            \vbox{\hbox{
$\displaystyle{10^3 \, {\Delta \Gamma \over \Gamma^{(0)}} }$
               \hskip -0.1in \null} \vskip 1.9in} &
            \epsfbox{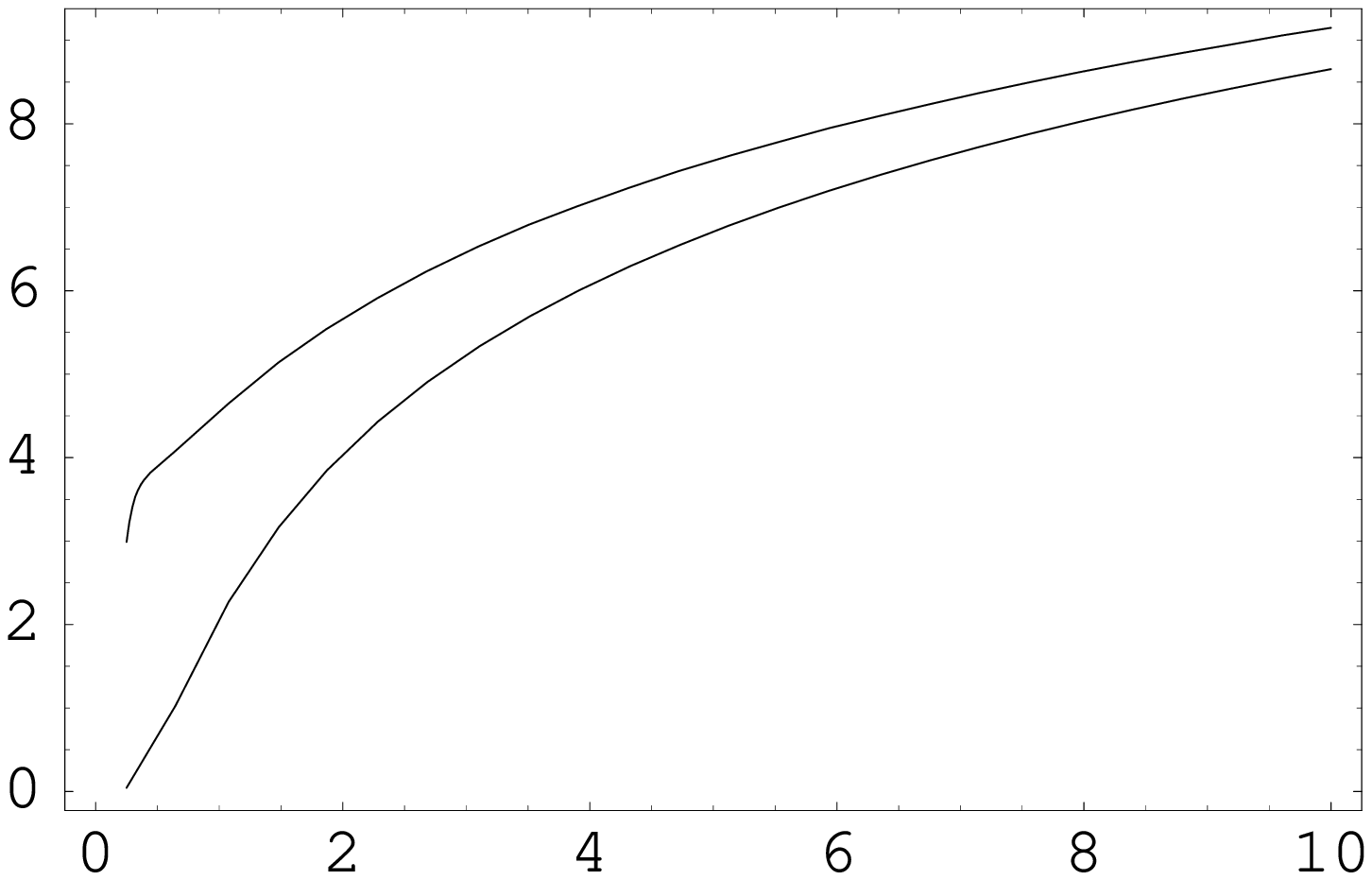} \\
            &
            \hbox{$T$  (MeV) } \\
        \end{tabular}
    \end{center}
\label{fig1}
\vskip 1in   
\protect\caption
	{%
The lower curve is ($10^3$ times) 
the ratio of the finite-temperature part of the
leading electromagnetic corrections to the neutron disappearance 
rate $\Delta \Gamma^{(T)}_n$, given in Eq.~(\ref{total}), to the free
neutron rate $\Gamma_n^{(0)}$, given in Eq.~(\ref{freeneutron}). The
upper curve is the same ratio for the proton.  
	}
\end {figure}

\newpage

\begin {figure}[ht]
    \begin{center}
        \epsfxsize 5.6in
        \begin{tabular}{rc}
            \vbox{\hbox{
$\displaystyle{10^3 \, {\Delta \Gamma \over \Gamma^{(0)}}} $
               \hskip -0.1in \null} \vskip 1.9in} &
            \epsfbox{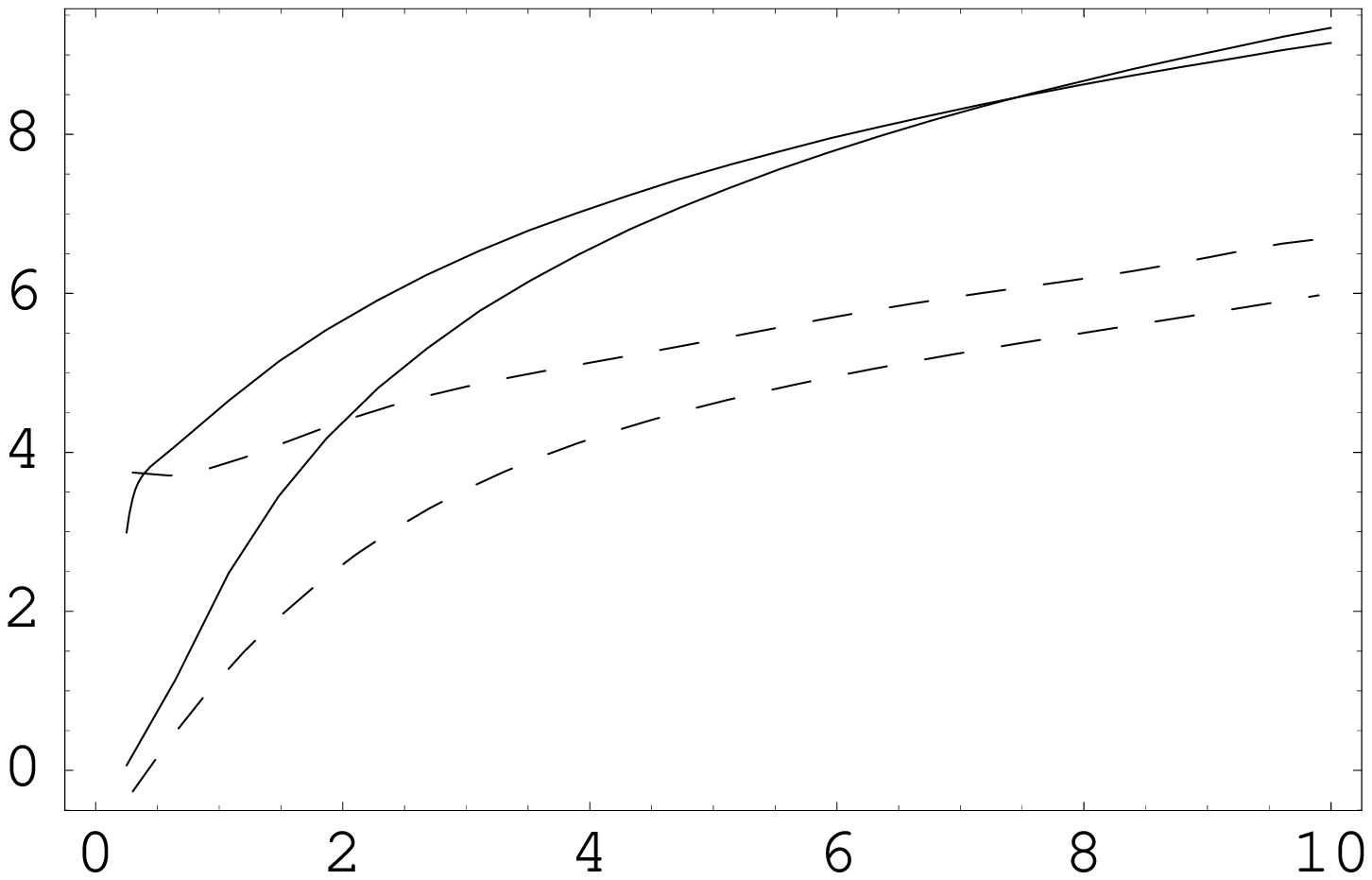} \\
            &
            \hbox{$T$  (MeV)} \\
        \end{tabular}
    \end{center}
\label{fig2}
\vskip 1in   
\protect\caption
	{%
Our results shown as solid lines compared to the results of Lopez and
Turner [6] shown as dashed lines. As in Fig.~1, the curves describe
the temperature-dependent part of the leading electromagnetic
corrections to the rates divided by the corresponding uncorrected
rates, $ \Delta \Gamma^{(T)} / \Gamma^{(0)}$. The upper curves refer
to the proton rates, the lower curves to the neutron rates. As
described in the text, we plot our ``revised'' result for the neutron
rate obtained by adding the contribution (\ref{repeatforgot}) so as to
account for the previously omitted process. To obtain the Lopez and
Turner comparison figures, we combined their results for the ``finite
temperature radiative corrections", as shown in their Fig. 11, with
their results for the ``finite temperature electron-mass correction",
as shown in their Fig. 16.
	}
\end {figure}

\newpage

\begin {figure}[ht]
    \begin{center}
        \epsfxsize 5.6in
        \begin{tabular}{rc}
            \vbox{\hbox{
$\displaystyle{ 10^4 \, {\Delta \Gamma \over \Gamma^{(0)}} } $
               \hskip -0.1in \null} \vskip 1.9in} &
            \epsfbox{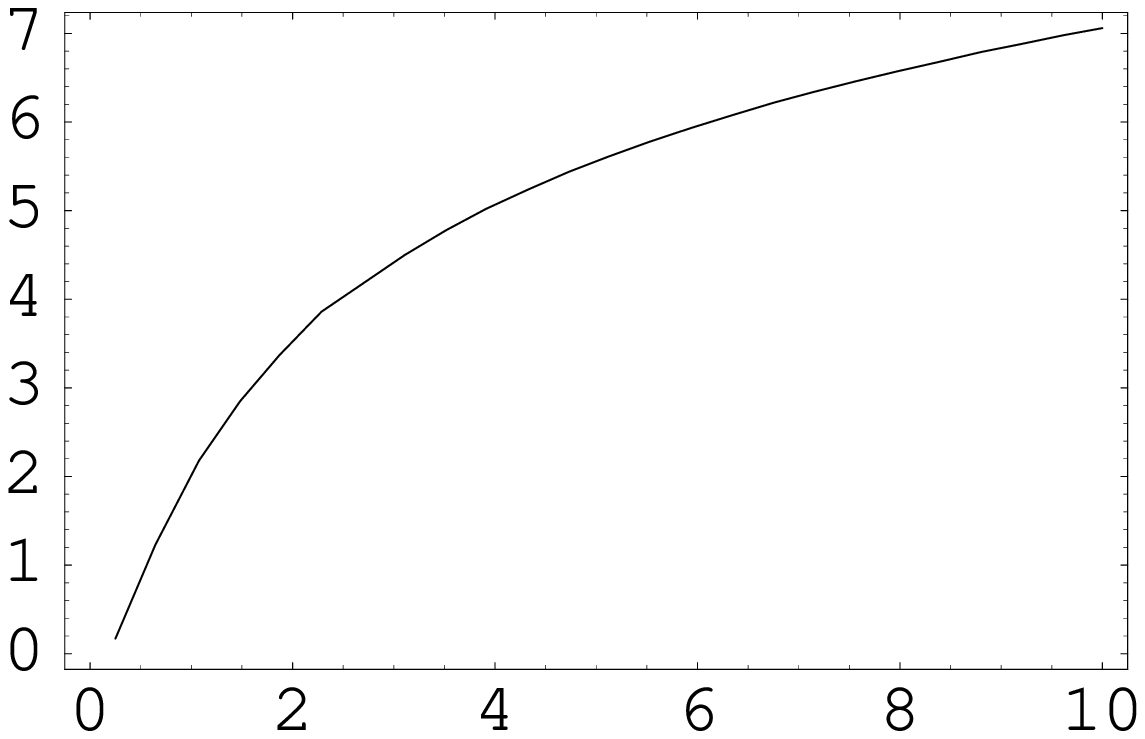} \\
            &
            \hbox{$T$  (MeV)} \\
        \end{tabular}
    \end{center}
\label{fig3}
\vskip 1in   
\protect\caption
	{%
Ratio of the previously omitted processes 
$ \nu + e^+ + n  \leftrightarrow  \gamma + p $
contribution to the neutron disappearance rate 
$\Delta \Gamma_{n \, {\rm F}} $, as given in Eq.~(\ref{fforgive}), 
to the uncorrected rate
$\Gamma_n^{(0)}$ given in Eq.~(\ref{freeneutron}), multiplied by $10^4$.  
Since the complete contributions of a new set of
processes obey detailed balance, the corresponding ratio for a proton
is the same. Note that the contribution of these processes is about an
order-of-magnitude smaller than the $\Delta \Gamma^{(T)} /
\Delta \Gamma^{(0)}$ ratios displayed in Fig.~2, and thus they cannot
account for the discrepancies appearing in Fig.~2.
	}
\end {figure}

\newpage

\appendix

\section{Rate Formula Derivation}

As discussed in the text, plasma modifications of the rates involve
low-energy, soft processes.  Thus they may be calculated using the
limit in which the nucleons are treated as being very heavy and 
described by operators $\Psi^\dagger({\bf r},t)$ and 
$\Psi({\bf r},t)$ that create and destroy a particle at the spatial
point ${\bf r}$ at time $t$. The weak interaction Hamiltonian may
be expressed as
\begin{equation}
H_W(t)  = {\cal K}(t) + {\cal K}^\dagger(t) \,,
\end{equation}
where 
\begin{eqnarray}
{\cal K}(t) &=& {G_W \over \sqrt 2} \int(d^3{\bf r}) 
\Bigr\{ g_V \, \psi^\dagger_\nu(x) \left( 1 - \gamma_5 \right)
\psi_e(x) \, \Psi^\dagger_n(x) \Psi_p(x) 
\nonumber\\
&& \qquad\qquad
+g_A \, \psi^\dagger_\nu(x)\alpha_l \left(1 - \gamma_5 \right) \psi_e(x)
\, \Psi^\dagger_n(x) \sigma_l \Psi_p(x)  \Bigr\} \,.
\label{local}
\end{eqnarray}
Here $\psi_e$ and $\psi_\nu$ are the Dirac fields for the electron and
(electron) neutrino, and $\alpha_l = \gamma^0 \gamma_l $ is the usual
Dirac matrix.

The operator 
\begin{equation}
N_\nu(t) = \int (d^3{\bf r}) \psi^\dagger_\nu(x) \psi_\nu(x)
\end{equation}
measures the leptonic charge of the neutrinos, the number of $\nu$
minus the number of $\bar\nu$.  The generic equal-time 
anticommutator for Fermi fields reads
\begin{equation}
\left\{ \psi({\bf r},t) \,, \psi^\dagger({\bf r}',t) \right\}
  = \delta({\bf r} - {\bf r}') \,,
\label{acomm}
\end{equation}
where the $\delta$ function on the right implicitly includes the Dirac
field indices. Using this anticommutator for the neutrino field gives
the time derivative
\begin{eqnarray}
\dot N_\nu(t) &=& i \left[ H_W(t) \,, N_\nu(t) \right]
\nonumber\\
&=& -i \left[ {\cal K}(t) - {\cal K}^\dagger(t) \right] \,.
\end{eqnarray}
The operators which appear here create and destroy electrons and
neutrinos, and hence their expectation vanishes in the unperturbed
plasma ensemble, which is diagonal in these particle numbers.  The
reaction rate $\Gamma_\nu$ appears in the additional linear 
response\footnote{We use here the general method advocated by Brown
and Sawyer \cite{BS}.} to the effect of the perturbation $H_W(t)$,
\begin{eqnarray}
\Gamma_\nu &=& -i\int_{- \infty}^0 dt \left\langle
\left[ \dot N_\nu(0) \,, H_W(t) \right] \right\rangle_T
\nonumber\\
&=&  - \int_{- \infty}^0 dt \left\langle \left[
	{\cal K}(0) - {\cal K}^\dagger(0) \,, 
	{\cal K}(t) + {\cal K}^\dagger(t) \right] \right\rangle_T \,.
\end{eqnarray}
Since ${\cal K}$ does not conserve particle number,
\begin{equation}
\left\langle\left[ {\cal K}(0) \,, {\cal K}(t)
\right]\right\rangle_T = 0 = \left\langle\left[
{\cal K}^\dagger(0) \,, {\cal K}^\dagger(t) \right]\right\rangle_T
\end{equation}
In virtue of the time transitional invariance of the ensemble,
\begin{equation}
\left\langle\left[ {\cal K}(0) \,, {\cal K}^\dagger(t)
\right]\right\rangle_T
= \left\langle\left[ {\cal K}(-t) \,, {\cal K}^\dagger(0) 
	\right]\right\rangle_T \,.
\end{equation}
Thus, we change the integration variable, $t\rightarrow
-t$, use the antisymmetry of the commutator, and combine terms combine
to obtain
\begin{equation}
\Gamma_\nu = - \int_{-\infty}^{\infty} dt 
\left\langle\left[ {\cal K}(t) \,, {\cal K}^\dagger(0) 
	\right]\right\rangle_T \,.
\end{equation}

Although the expectation value that appears here describes a thermal
ensemble of electrons and neutrinos, both of which are taken to have
zero chemical potentials, we take the expectation value to have
only a single nucleon, which we treat as a heavy particle located at
the coordinate origin, ${\bf r} = 0$.  Thus the nucleon destruction
operators in the weak Hamiltonian give a vanishing contribution 
unless their coordinate ${\bf r} = 0$, and so the spatial coordinates
of the electron and neutrino fields can be taken to vanish in the
local interaction (\ref{local}). 
Rotational invariance of the thermal ensemble implies that the
(time-component) vector -- (space-component) 
axial-vector interference terms vanish and that there can be no
dependence upon the nucleon spin orientation. Hence the result is
unchanged by averaging over the initial nucleon spin as well as
performing the required sum over the final nucleon spin.  Thus, we may
replace the (direct product) $\sigma_l \, \sigma_m $ in the
axial-vector -- axial-vector contribution by $\delta_{lm}$. Moreover,
as one can verify in detail from the structure of the neutrino field
thermal expectation values that will soon be presented, one can also
replace the resulting (direct product) $\alpha_l \, \alpha_l $ by
$3$. With these replacements made, both the vector and the
axial-vector parts of the weak Hamiltonian involve the isospin
lowering operator
\begin{equation}
T_-(t) = \int (d^3{\bf r}) \Psi^\dagger_n(x) \Psi_p(x) \,,
\end{equation}
and its Hermitian adjoint, the isospin raising operator
\begin{equation}
T_+(t) = T_-^\dagger(t) = 
	\int (d^3{\bf r}) \Psi^\dagger_p(x) \Psi_n(x) \,,
\end{equation}
with
\begin{eqnarray}
\Gamma_\nu &=& - {1\over2} G_W^2 \left( g_V^2 + 3 g_A^2 \right)
\int_{-\infty}^{+\infty} dt
\nonumber\\
&& \qquad \left\langle \left[ 
\psi_\nu^\dagger({\bf 0},t) \left( 1 - \gamma_5 \right) 
\psi_e({\bf 0},t) T_-(t) \,,
\psi_e^\dagger(0) \left( 1 - \gamma_5 \right) 
\psi_\nu(0) T_+(0) \right] \right\rangle_T  \,.
\label{rate}
\end{eqnarray}

We define the real-time, generic free Fermi field thermal expectation
values with vanishing chemical potential as
\begin{equation}
\langle \psi_\alpha(x) \psi_\beta^\dagger(x') \rangle^{(0)}_T
	= \left[ S^{(+)}(x-x') \gamma^0 \right]_{\alpha \beta} \,,
\label{poscorr}
\end{equation}
and
\begin{equation}
\langle \psi_\beta^\dagger(x') \psi_\alpha(x) \rangle^{(0)}_T
	= \left[ S^{(-)}(x-x') \gamma^0 \right]_{\alpha \beta} \,.
\end{equation}
These function satisfy the free Dirac equation.  Since these thermal
expectation values involve ${\rm Tr} e^{-\beta H} \cdots $, with
\begin{equation}
e^{-\beta H} \psi({\bf r}, t) =  \psi({\bf r}, t + i \beta) 
e^{-\beta H} \,,
\end{equation}
the cyclic symmetry of the trace provides the boundary condition
\begin{equation}
\left[ S^{(+)}({\bf r} - {\bf r}', t - t') \gamma^0 \right]_{\alpha \beta} =
\left[ S^{(-)}({\bf r} - {\bf r}', t + i \beta - t') 
\gamma^0 \right]_{\alpha \beta} \,.
\end{equation}
This boundary condition, plus that provided by the equal-time
anticommutation relation (\ref{acomm}), determines the solution of the free
Dirac equation to be
\begin{eqnarray}
S^{(\pm)}(x-x') \gamma^0 &=& \int { (d^3{\bf p}) \over (2\pi)^3}
{ 1 \over 2 E(p) } e^{i {\bf p} \cdot ( {\bf r} - {\bf r}' ) }
\nonumber\\
&& \quad 
\Bigg\{ \left[ E(p) + \gamma^0 m + \alpha \cdot {\bf p} \right]
	n\left(\mp E(p) \right) e^{-i E(p) (t - t') } 
\nonumber\\
&& \qquad
+ \left[ E(p) - \gamma^0 m - \alpha \cdot {\bf p}  \right]
	n\left(\pm E(p) \right) e^{+i E(p) (t - t') } \Bigg\} \,.
\label{corr}
\end{eqnarray}
Here
\begin{equation}
E(p) = \sqrt{ p^2 + m^2 } \,,
\end{equation}
and 
\begin{equation}
n(E) = { 1 \over e^{\beta E} + 1 } 
\end{equation}
is the Fermi thermal distribution appropriate for an initial state,
while
\begin{equation}
n(-E) = { e^{\beta E} \over e^{\beta E} + 1 } = 1 - n(E)
\end{equation}
gives the Pauli blocking factor appropriate for final states.

Since the rate (\ref{rate}) already has the weak interaction taken
into account, we may use the free field correlations (\ref{corr})
with $m=0$ for the neutrino fields.  Since the electron thermal
ensemble is parity invariant, a single factor of $\gamma_5$ does not
contribute while $\gamma_5^2 = 1$. Integrating over the neutrino solid
angles and writing $p = E_\nu$ for the massless neutrino, we get
\begin{eqnarray}
\Gamma_\nu &=& -  G_W^2 \left( g_V^2 + 3 g_A^2 \right)
{ 1 \over (2\pi)^2} \int_{-\infty}^{+\infty} dE_\nu \, E_\nu^2 \,
n(E_\nu) \,  \int_{-\infty}^{+\infty} dt
\nonumber\\
&&  \left\langle  
e^{+i E_\nu t} \psi_e({\bf 0},t) T_-(t) 
\psi_e^\dagger(0) T_+(0) -
e^{ -i E_\nu t} 
\psi_e^\dagger(0) T_+(0) 
 \psi_e({\bf 0},t) T_-(t) \right\rangle_T  \,.
\end{eqnarray}
The first term in the thermal expectation value only contributes when
the initial nucleon state is a neutron which is changed into a proton
by the action of $T_+(0)$; this term corresponds to the process $ \nu
+ n \to p + e^-$ plus the other processes related by crossing the
leptons and/or including photon emission.  These processes decrease
the number $\nu$ minus $\bar \nu$ in accord with the overall minus sign
out in front.  Thus the positive rate $\Gamma_n$ for the disappearance
of an initial neutron is given by
\begin{eqnarray}
\Gamma_n &=&   G_W^2 \left( g_V^2 + 3 g_A^2 \right)
{ 1 \over (2\pi)^2} 
\nonumber\\
&& \qquad
\int_{-\infty}^{+\infty} dE_\nu \, E_\nu^2 \,
n(E_\nu) \,  \int_{-\infty}^{+\infty} dt
\,  \left\langle  
e^{+i E_\nu t} \psi_e({\bf 0},t) T_-(t) 
\psi_e^\dagger(0) T_+(0) \, \right\rangle_T  \,.
\label{n-to-p}
\end{eqnarray}
Similarly, the rate for the disappearance of an initial proton is
given by
\begin{eqnarray}
\Gamma_p &=&   G_W^2 \left( g_V^2 + 3 g_A^2 \right)
{ 1 \over (2\pi)^2} 
\nonumber\\
&& \qquad 
\int_{-\infty}^{+\infty} dE_\nu \, E_\nu^2 \,
n(E_\nu) \,  \int_{-\infty}^{+\infty} dt
 \left\langle  e^{ -i E_\nu t} 
\psi_e^\dagger(0) T_+(0) 
 \psi_e({\bf 0},t) T_-(t) \right\rangle_T  \,.
\end{eqnarray}

\section{Electron Self-Energy Contributions}

The electron self-energy function for the thermal electron Green's
function, is given, to  one-loop order, by
\begin{equation}
\Sigma(p) = - T \sum_{n=-\infty}^{n=+\infty} \int { (d^3{\bf k}) \over
(2\pi)^3 } \gamma_\mu { m - \gamma(p+k) \over (p+k)^2 + m^2 }
\gamma_\nu  \, e^2 \, D_{\mu\nu}(k) \,.
\label{f1}
\end{equation}
The four vectors here are in Euclidean space with the scalar product 
$\gamma(p+k) = \gamma \cdot ({\bf p}+{\bf k}) + \gamma_4 \,(p+k)_4 $,
$ \gamma_4 = -i \gamma^0 $, and $k_4 = \omega_n = 2\pi \, n \, T$ while
the external energy takes on the values $p_4 = (2m + 1 ) \pi \,
T$. The radiation gauge is chosen where
\begin{equation}
D_{lm}(k) = \left[ \delta_{lm} - { k_l k_m \over {\bf k}^2 }
\right]  { k \over k^2} \,,
\label{f2} 
\end{equation}
and
\begin{equation}
 D_{44}(k) = { 1 \over {\bf k}^2 } \,.
\label{f3}
\end{equation}

The frequency sum is conveniently performed by expressing it as a contour
integral, 
\begin{equation}
T \, \sum_{n=-\infty}^{n=+\infty} \, \cdots =
\int_C { d \omega \over 2\pi i} \, {1 \over 2}
\cot\left({\beta\omega \over 2} \right) \, \cdots \,,
\label{f4}
\end{equation}
where the contour $C$ runs from $-\infty$ to $+\infty$ just below the 
real axis and then returns to $-\infty$ just above the real axis.  For
the part of the contour just below the real axis, we write
\begin{equation}
{1 \over 2i} \cot\left({\beta\omega \over 2} \right) =
{ 1 \over 2} + { 1 \over e^{i\beta\omega} - 1} \,,
\label{f5}
\end{equation}
with the second term exponentially damped in the lower-half plane
where we shall close the contour for it. For the upper-half portion of
the contour, we write
\begin{equation}
{1 \over 2i} \cot\left({\beta\omega \over 2} \right) =
- { 1 \over 2} - { 1 \over e^{-i\beta\omega} - 1} \,,
\label{f6}
\end{equation}
and close the contour for the second term in the upper-half plane
where it is exponentially damped. In this way, we achieve the
separation
\begin{equation}
\Sigma(p) = \Sigma^{(0)}(p) + \Sigma^{(T)}(p) \,,
\label{f7}
\end{equation}
where 
\begin{equation}
\Sigma^{(0)}(p) = - e^2\int{ d\omega \over 2\pi} \int { (d^3{\bf k}) \over
(2\pi)^3 } \gamma_\mu { m - \gamma(p+k) \over (p+k)^2 + m^2 }
\gamma_\nu  \, D_{\mu\nu}(k) 
\label{f8}
\end{equation}
is the vacuum self-energy function (in Euclidean space).  This vacuum
part must be regulated and renormalized. The remaining,
temperature-dependent piece $\Sigma^{(T)}(p)$ is a finite,
well-defined function which vanishes when the temperature
vanishes.

In closing the contours for the thermal part of the self energy, poles
are encountered at $ \omega = \pm i k$, where $k = |{\bf k}|$,
producing a factor of the Bose distribution
\begin{equation}
f(k) = { 1 \over e^{\beta k} - 1 } \,.
\end{equation}
Poles are also encountered at 
$\omega = \pm i \, E({\bf p}+{\bf k}) - p_4 $, 
where $ E({\bf p}+{\bf k}) = \sqrt{({\bf p} + {\bf k})^2 + m^2} $. 
Since $\beta p_4 = (2m + 1 ) \pi$, with
$\exp\{ i \beta p_4 \} = -1$, these poles are accompanied by a factor
of the Fermi distribution
\begin{equation}
n(E({\bf p}+{\bf k}) ) = { 1 \over e^{\beta E({\bf p}+{\bf k}) } + 1 } \,.
\end{equation}
Separating out the Coulomb and radiative pieces, and continuing to 
real energy by setting  $p_4=-iE$, we obtain
\begin{equation}
\Sigma^{(T)}(p) = \Sigma^{(T)}_R(p) + \Sigma_C^{(T)}(p) \,,
\end{equation}
where
\begin{eqnarray}
\tab \Sigma^{(T)}_R({\bf p},E) = - e^2\int { (d^3{\bf k}) \over (2\pi)^3} 
 \, \left( \delta_{lm} - \hat k_l \hat k_m \right)
\nonumber\\
\tab \qquad\qquad
\gamma_l  \, \Bigg\{ {1 \over 2 k} f(k)
\Bigg[ { m - \gamma \cdot ({\bf p} + {\bf k} ) + \gamma^0 
( E + k ) \over E({\bf p}+{\bf k})^2 - (E + k)^2  }
+ { m - \gamma \cdot ({\bf p} + {\bf k} ) +\gamma^0 
(E-k) \over E({\bf p}+{\bf k})^2 - ( E - k )^2} \Bigg]
\nonumber\\
\tab \qquad\qquad\qquad
- {1 \over 2 E({\bf p}+{\bf k}) } \, n(E({\bf p}+{\bf k}))
\Bigg[{ m - \gamma \cdot ({\bf p} + {\bf k} ) +\gamma^0 
 E({\bf p}+{\bf k}) \over k^2 - (E - E({\bf p}+{\bf k}) )^2 } 
\nonumber\\
\tab \qquad\qquad\qquad\qquad
+
{ m - \gamma \cdot ({\bf p} + {\bf k} ) - \gamma^0 
E({\bf p}+{\bf k}) \over k^2 - (E + E({\bf p}+{\bf k}) )^2 } 
\Bigg] \Bigg\}  \gamma_m \,, 
\label{f12}
\end{eqnarray}
and
\begin{equation}
\Sigma^{(T)}_C({\bf p},E) = - 2 e^2 \int { (d^3{\bf k}) \over (2\pi)^3} 
{1 \over 2 E({\bf p}+{\bf k}) } \, n(E({\bf p}+{\bf k}) ) \, 
\left[ { m + \gamma \cdot ({\bf p} + {\bf k} )\over k^2} 
 \right] \,.
\label{f13}
\end{equation}

A partial fraction decomposition in the radiative piece produces a
convenient alternative form: 
\begin{eqnarray}
\Sigma^{(T)}_R({\bf p},E) &=& - e^2\int { (d^3{\bf k}) \over (2\pi)^3} 
 \, \left( \delta_{lm} - \hat k_l \hat k_m \right)
{ 1 \over 4 \,  k \, E({\bf p}+{\bf k}) }
\nonumber\\
&&
\gamma_l  \, \Bigg\{ 
\left[  m - \gamma \cdot ({\bf p} + {\bf k} ) + \gamma^0 
E({\bf p}+{\bf k}) \right] 
\nonumber\\
&& \qquad
\left[ { f(k) + n(E({\bf p}+{\bf k})) \over 
E({\bf p}+{\bf k}) - E - k } + { f(k) - n(E({\bf p}+{\bf k})) \over
E({\bf p}+{\bf k}) - E + k } \right]
\nonumber\\
&& \quad +
\left[  m - \gamma \cdot ({\bf p} + {\bf k} ) - \gamma^0 
E({\bf p}+{\bf k}) \right] 
\nonumber\\
&& \qquad
\left[ { f(k) - n(E({\bf p}+{\bf k})) \over 
E({\bf p}+{\bf k}) + E + k } + { f(k) + n(E({\bf p}+{\bf k})) \over
E({\bf p}+{\bf k}) + E - k } \right] \Bigg\} \gamma_m \,.
\label{fract}
\end{eqnarray}
Thus we find that the gamma-matrix trace in the basic rate formula 
(\ref{wonder}) involves, for the radiative part,
\begin{eqnarray}
r_\pm &=& {1 \over 4} \left( \delta_{lm} - \hat k_l \hat k_m \right) 
{\rm tr} \, (m - \gamma p ) \gamma^0 (m -\gamma p)
\nonumber\\
&& \qquad\qquad
\gamma_l \left[  m - \gamma \cdot ({\bf p} + {\bf k} ) \pm \gamma^0 
E({\bf p}+{\bf k}) \right] \gamma_m 
\nonumber\\
&=&
-4 \, E \, m^2 + E \, k^2 - (E/ k^2)  \, \left[ E^2({\bf p} + {\bf k} )
	- E^2({\bf p}) \right]^2  
\nonumber\\
&& \qquad\qquad\qquad
\pm 2 \, E({\bf p}+{\bf k}) 
\left[ E^2 + E^2({\bf p}) \right] \,.
\label{tracer}
\end{eqnarray}
The trace for the Coulomb part is given by
\begin{eqnarray}
c &=& {1 \over 4} \, {\rm tr} \, (m - \gamma p ) \gamma^0 (m -\gamma p)
\left[  m + \gamma \cdot ({\bf p} + {\bf k} ) \right] 
\nonumber\\
&=& 
2 \, E \, \left[ E^2({\bf p}) + {\bf p} \cdot {\bf k} \, \right] \,.
\label{coulbb}
\end{eqnarray}

We begin by calculating the ${\rm Im} \, \Sigma$ contribution in the
rate formula (\ref{wonder}), which we shall call 
$\Gamma_{n \, {\rm e-e}}^{(\gamma)} $. The complete self-energy function 
(with no $T=0$ subtraction) enters here. Returning for a moment to
recall the discussion of our evaluation of the frequency sum 
(\ref{f4}), it is easy to see that the complete $\Sigma$ function 
can be obtained from the result (\ref{fract}) for $\Sigma^{(T)}$
by making the replacements, $f(k)\to [f(k)+1/2]$ and
$n(E({\bf p} + {\bf k})) \to [ n(E({\bf p} + {\bf k})) - 1/2 ]$. 
We incorporate this replacement and insert the imaginary part of the
result (\ref{fract}) evaluated with the trace formula (\ref{tracer}) 
into the basic rate formula (\ref{wonder}). We  interchange the
${\bf p}$ and ${\bf k}$ integrals and then replace the ${\bf p}$
integration variable by ${\bf q} = {\bf p} + {\bf k}  $. In this way,
we obtain
\begin{eqnarray}
\Gamma_{n \, {\rm e-e}}^{(\gamma)} &=& 
 {e^2  G_W^2 \over (2\pi)^2 } \left( g_V^2 + 3 g_A^2 \right)
 \int_{-\infty}^{+\infty} dE \, \chi(E)  
\nonumber\\
&& 
\int { (d^3{\bf k}) \over (2\pi)^3} \, {1 \over k}
\, \int { (d^3{\bf q}) \over (2\pi)^3} \, {1 \over E({\bf q}) } 
\, { 2 \pi \over [ E^2({\bf q} - {\bf k}) - E^2 ]^2 }
\nonumber\\
&&
\Bigg\{ \left[ f(k) + n(E({\bf q})) \right] 
\left[ r_+ \, \delta\left( E + k -  E({\bf q}) \right) -
	r_- \, \delta\left( E - k +  E({\bf q}) \right) \right]
\nonumber\\
&+& 
 \left[ f(k) - n(E({\bf q})) + 1 \right] 
\left[ r_+ \, \delta\left(E - k -  E({\bf q}) \right) -
	r_- \, \delta\left(E + k +  E({\bf q}) \right) \right]
\Bigg\} \,.
\nonumber\\
&&
\label{mess}
\end{eqnarray}
The only angular dependence that appears here is in 
\begin{equation}
E({\bf q} - {\bf k})^2 = E({\bf q}) ^2+ k^2 + {\bf q} \cdot {\bf k} \,,
\end{equation}
which enters in the denominator and in the definition (\ref{tracer}) of
$r_\pm$. Taking ${\bf k}$ as the $z$ axis and performing the solid
angle integral for ${\bf q}$ and then the remaining solid angle 
integral for ${\bf k}$, and using the delta functions to
remove the $E$ integral gives, after one final variable change to
$ {\cal E} = \sqrt{ q^2 + m^2 } $
, and some algebraic effort, 
\begin{eqnarray}
\Gamma_{n \, {\rm e-e}}^{(\gamma)} &=& 
2\, {e^2  G_W^2 \over (2\pi)^5 } \left( g_V^2 + 3 g_A^2 \right)
 \int_0^\infty {dk \over k}  \, \int_m^\infty d{\cal E}  
\nonumber\\
&&
\Bigg\{ \left[ f(k) + n({\cal E}) \right] 
\left[ \chi({\cal E} - k) +  \chi(- {\cal E} + k) \right] F_-
\nonumber\\
&+& 
 \left[ f(k) - n({\cal E}) + 1 \right] 
\left[ \chi({\cal E} + k) +  \chi(- {\cal E} - k) \right] F_+
\Bigg\} \,,
\label{gulpp}
\end{eqnarray}
where
\begin{equation}
F_\pm = A \pm k \, B \,,
\label{fpm}
\end{equation}
with
\begin{equation}
A = 
\left[ 2 \, {\cal E}^2 + k^2 \right]
	\ln\left({ {\cal E} + q \over {\cal E} - q } \right)
		- 4 \, q \,  {\cal E}  \,,
\label{aaa}
\end{equation}
and
\begin{equation}
B =  2 \, {\cal E} \, 
	\ln\left({ {\cal E} + q \over {\cal E} - q } \right)
		- 4 \, q  \,.
\label{bbb}
\end{equation}
We have also made use the function defined in 
Eq.~(\ref{chidef}), which we repeat here for convenience, 
\begin{equation} 
\chi(E) = n(E-\Delta) n(-E) (E-\Delta)^2 \,.
\label{repeat}
\end{equation} 

To facilitate the removal of the ``$T=0$'' piece and to also put the
result in a form that clarifies its structure, we note that
\begin{eqnarray}
f(k) + n({\cal E}) &=& f(k) \, n(-{\cal E}) \, n(- {\cal E} + k )^{-1} 
\nonumber\\
&=& -f(-k) \, n( {\cal E}) \, n( {\cal E} - k )^{-1} \,,
\end{eqnarray}
and
\begin{eqnarray}
f(k) - n({\cal E}) + 1 &=& -f(-k) \, 
n(-{\cal E}) \, n(- {\cal E} - k )^{-1} 
\nonumber\\
&=& f(k) \, n( {\cal E}) \, n( {\cal E} + k )^{-1} \,.
\end{eqnarray}
Here
\begin{equation}
-f(-k) = 1 + f(k) 
\end{equation}
and
\begin{equation}
n(-{\cal E}) = 1 - n({\cal E})
\end{equation}
describe the occupation factors appropriate for final states: 
$ 1 + f(k) $ gives the Bose enhancement factor and $ 1 - n({\cal E})$
the Pauli blocking factor. Using these identities, we obtain
\begin{eqnarray}
\Gamma_{n \, {\rm e-e}}^{(\gamma)} &=& 
2\, {e^2  G_W^2 \over (2\pi)^5 } \left( g_V^2 + 3 g_A^2 \right)
 \int_0^\infty {dk \over k}  \, \int_m^\infty d{\cal E}  
\nonumber\\
&&
\Bigg\{
\left[ f(k)\,  n(- {\cal E}) \, \tilde\chi({\cal E} - k) 
- f(-k) \,  n({\cal E}) \,  \tilde\chi(- {\cal E} + k) \right] F_-
\nonumber\\
&& \quad + 
\left[ -f(-k) \, n(-{\cal E}) \, \tilde\chi({\cal E} + k) 
+ f(k) \,  n({\cal E}) \, \tilde\chi(- {\cal E} - k) \right] F_+
\Bigg\} \,,
\label{ggulpp}
\end{eqnarray}
where
\begin{equation}
\tilde \chi(E) =  n(E-\Delta) \, (E-\Delta)^2 \,.
\label{tildechi}
\end{equation}
The energy integration variable $E$ in Eq.~(\ref{mess}) differs
from the neutrino energy by a positive constant, 
$E = E_\nu + \Delta$. 
Neutrinos in the initial state are described by positive
values of the neutrino energy, $E_\nu > 0$.  Thus the terms in
Eq.~(\ref{mess}) involving  
$ \delta( E - k \, + \cdots )$, 
where 
$k = |{\bf k}|$ 
is an on mass shell, real photon energy, correspond to the
production of photons in the final state. These real photon
contributions appear in Eq.~(\ref{ggulpp}) in the terms involving 
$\tilde\chi(-{\cal E} + k)$ and $\tilde\chi({\cal E} + k)$. These are
just the terms that have the occupancy factors $f(-k)$ that are
appropriate for produced photons. It is a simple matter to check that
the Fermi factors $n(\pm E)$ are just those needed in the various
photon absorption and emission processes described by
Eq.~(\ref{ggulpp}), but we shall not bother to enumerate then
here. We just note that the ``$T=0$'' part of the rate is obtained by
setting $f(k) = 0$. Subtracting this part is equivalent to the
substitution $-f(-k) \to f(k)$, and so the $T\ne 0$, photon thermal 
bath contribution is given by
\begin{eqnarray}
\Gamma_{n \, {\rm e-e}}^{(\gamma,\gamma)} &=& 
2\, {e^2  G_W^2 \over (2\pi)^5 } \left( g_V^2 + 3 g_A^2 \right)
 \int_0^\infty {dk \over k}  \, f(k) \, \int_m^\infty d{\cal E}  
\nonumber\\
&& 
\Big[
\left[ n(-{\cal E}) \tilde\chi({\cal E} - k) + 
 n({\cal E}) \tilde \chi(- {\cal E} + k) \right] F_-
+ \left[ n(-{\cal E}) \tilde\chi({\cal E} + k) +  
	 n({\cal E}) \tilde\chi(- {\cal E} - k) \right] F_+
	\Big] \,.
\nonumber\\
&&
\label{yetinfr}
\end{eqnarray}
Although this set of terms has no ultraviolet divergences, 
since $f(k) \, k^{-1} \to
k^{-2}$ as $k \to 0$, it is infrared divergent.  This contribution is
rendered finite in the infrared by other contributions that we are 
about to compute.  

We pause for a moment from our development to identify the
contribution from the rate
$e^+ + \nu + n \to \gamma + p$
which has been omitted from all previous papers.
Since it involves real photon production, it must involve
the terms containing $-f(-k)$ in
Eq.~(\ref{ggulpp}).  The second term involving the
$\tilde\chi({\cal E} + k)$ function describes an incident
neutrino of energy $ E_\nu = {\cal E} + k - \Delta $ corresponding to
a produced electron of energy ${\cal E}$ rather than an initial
positron of this energy. Hence it is the first term containing $-f(-k)$
that contributes to the process in question, with the photon energy
starting at $k = {\cal E} + \Delta$ corresponding to an incident
neutrino with zero energy. Thus 
the previously forgotten contribution to the rate is given by
\begin{eqnarray}
\Gamma_{n \, {\rm F}} &=& 
2\, {e^2  G_W^2 \over (2\pi)^5 } \left( g_V^2 + 3 g_A^2 \right)
\, \int_m^\infty d{\cal E}  \,  n({\cal E}) 
 \int_{{\cal E} + \Delta}^\infty {dk \over k} \, 
\left[ - f(-k) \right] \,
\tilde\chi(-{\cal E} + k ) \, F_- \,.
\label{forgive}
\end{eqnarray}
The $T=0$ part of this rate is obtained by taking $-f(-k) \to 1$,
\begin{eqnarray}
\Gamma_{n \, {\rm F}}^{(T=0)} &=& 
2\, {e^2  G_W^2 \over (2\pi)^5 } \left( g_V^2 + 3 g_A^2 \right)
\, \int_m^\infty d{\cal E}  \,  n({\cal E}) 
 \int_{{\cal E} + \Delta}^\infty {dk \over k} \, 
\tilde\chi(-{\cal E} + k ) \, F_- \,.
\label{forgot}
\end{eqnarray}

We turn now to evaluate the on mass shell $ E = \pm E({\bf p}) $
contributions that appear in the last two lines of the rate formula
(\ref{wonder}). This piece of the rate formula involves the
temperature-dependent self-energy function in the form
\begin{equation}
\bar\Sigma^{(T)}({\bf p},E) =
{\rm tr} \, (m - \gamma p) \, \gamma^0 \, (m - \gamma p) \,
\Sigma^{(T)}({\bf p},E) \,.
\end{equation}
We separate out the parts that involve the photon and electron
phase-space densities by writing 
\begin{equation}
\bar\Sigma^{(T)}({\bf p},E) = \bar\Sigma^{(T)}_\gamma({\bf p},E) +
\bar\Sigma^{(T)}_e({\bf p},E) \,.
\label{decomp}
\end{equation}
We use of the previous results (\ref{f13}) -- (\ref{coulbb}) and a
little algebra to write the two parts as
\begin{eqnarray}
\bar\Sigma^{(T)}_\gamma({\bf p} ,E) &=&  4
e^2 \int { (d^3{\bf k}) \over (2\pi)^3} \, { 1 \over  k  } \, f(k) 
\Bigg\{2 E \Bigg[ m^2 + {\bf p}\cdot{\bf k} +
{ ( {\bf p}\cdot{\bf k} )^2 \over k^2}  \Bigg]
\nonumber\\
&& \qquad\qquad
\Bigg[ {1 \over E^2({\bf q}) - (E + k)^2 } 
	+  {1 \over E^2({\bf q}) - (E - k)^2 } \Bigg]
\nonumber\\
&-& 
  \left[ E^2 + E^2({\bf p}) \right] 
\Bigg[ { E + k \over E^2({\bf q}) - (E + k)^2 } 
	+  { E - k \over E^2({\bf q}) - (E - k)^2 } \Bigg]
\Bigg\} ,
\nonumber\\
&&
\label{gammm}
\end{eqnarray}
and
\begin{eqnarray}
\bar\Sigma^{(T)}_e({\bf p} ,E) &=& - 2
e^2 \int { (d^3{\bf q}) \over (2\pi)^3} \, { 1 \over  E({\bf q}) } 
\, n(E({\bf q})) 
\Bigg\{ \, { 4 E \over k^2 } 
		\, \left[ E^2({\bf p}) + {\bf p}\cdot{\bf k} \right]
\nonumber\\
&& \qquad
  + E\, \Bigg[  k^2 - 4m^2  - { 1 \over k^2} 
\left[ E^2({\bf q}) - E^2({\bf p}) \right]^2 \Bigg]
\nonumber\\
&& \qquad\qquad
\Bigg[ {1 \over (E({\bf q}) - E)^2  - k^2 } 
	+  {1 \over (E({\bf q}) + E)^2 - k^2 } \Bigg]
\nonumber\\
&+& 
 2 E({\bf q}) \, \left[ E^2 + E^2({\bf p}) \right] 
\Bigg[ {1 \over (E({\bf q}) - E)^2  - k^2 } 
	-  {1 \over (E({\bf q}) + E)^2 - k^2 } \Bigg]
\Bigg\} .
\nonumber\\
&&
\label{eee}
\end{eqnarray}
Here we have again introduced the variable 
${\bf q} = {\bf p} + {\bf k} $
and used it as the integration variable for the electron contribution
to the self energy (\ref{eee}). We have used the different integration
variables ${\bf k}$ and ${\bf q}$ for the photon and electron 
contributions because then the angular integrations appear in a simple
form. The terms on the first line on the right-hand side of
Eq.~(\ref{eee}) come from the Coulomb contribution (\ref{f13}) to the
self energy. 

To proceed with our computation, we note that the self-energy
function evaluated on the mass shell is gauge invariant\footnote{That
is, the result would not be altered if we changed our use of the
radiation (or Coulomb) gauge to a relativistic gauge such as Landau of
Feynman gauge.} and gives an energy shift.\footnote{See, for example,
the discussion of Eq.~(8.7.13) in Section 8 of \cite{brown}.} For
positive energies (particles) with spinors that obey
\begin{equation}
{\sum}_\lambda u_\lambda(p) \, \bar u_\lambda(p) = m - \gamma p \,,
\end{equation}
we have
\begin{equation}
2 E \, \Delta E = \bar u_\lambda(p) \, \Sigma \, u_\lambda(p)
\,,
\end{equation}
while for negative energies (antiparticles) with spinors 
\begin{equation}
{\sum}_\lambda v_\lambda(p) \, \bar v_\lambda(p) = m + \gamma p \,,
\end{equation}
the energy shift is
\begin{equation}
2 E \, \Delta E = \bar v_\lambda(p) \, \Sigma \, v_\lambda(p)
\,,
\end{equation}
Thus, in either case, one of the pieces of the rate formula
(\ref{wonder}) may be expressed as
\begin{eqnarray}
8 \, E^2 \, \Delta E^{(T)}({\bf p}) &=&  
{\rm tr} \, \gamma^0 \, (m - \gamma p) \,
\Sigma^{(T)}({\bf p},E) \, ( m - \gamma p) \Bigg|_{E = \pm E({\bf p})}
\nonumber\\
&=& 
\bar\Sigma^{(T)}({\bf p},E) \Bigg|_{E = \pm E({\bf p})}
\label{eshift}
\end{eqnarray}
Accordingly, we write the last, on-mass-shell contribution to the 
rate formula (\ref{wonder}) as
$ 
\Gamma_{n \, {\rm e-e}}^{(\Delta E)} + 
	\Gamma_{n \, {\rm e-e}}^{(Z)} 
$,
where 
\begin{eqnarray}
\Gamma_{n \, {\rm e-e}}^{(\Delta E)} &=& 
G_W^2  \left( g_V^2 + 3 g_A^2 \right) { 1 \over \pi} 
\int_{-\infty}^{+\infty} dE \, \int { (d^3{\bf p}) \over (2\pi)^3} \,
\nonumber\\
&& \qquad
\left[ \delta \left( E - E({\bf p}) \right) +  
\delta \left( E + E({\bf p}) \right) \right] \,
\Delta E^{(T)}({\bf p}) \,
{ \partial \over \partial E } \,  \chi(E)  \,,
\label{shift}
\end{eqnarray}
and
\begin{eqnarray}
\Gamma_{n \, {\rm e-e}}^{(Z)} &=& 
{ G_W^2 \over (2\pi)^2 } \left( g_V^2 + 3 g_A^2 \right)
\int_{-\infty}^{+\infty} dE \, \int { (d^3{\bf p}) \over (2\pi)^3} \,
 { \pi \over 2 E({\bf p} ) } 
\nonumber\\
&& \quad
 \left[ \delta \left( E - E({\bf p}) \right) -  
\delta \left( E + E({\bf p}) \right) \right]
\, \chi(E) \,
 { \partial \over \partial E } \, \left[ E^{-1} \, 
\bar\Sigma^{(T)}({\bf p}, E ) \right] \,.
\label{wave}
\end{eqnarray}
Referring back to the form of the uncorrected neutron rate 
(\ref{freeneutron}), we see that Eq.~(\ref{shift}) describes precisely
the correction to that rate which results from the (gauge-invariant)
energy shift $\Delta E^{(T)}({\bf p})$. As we shall
see, this temperature-dependent correction is well defined with no
infrared divergences.  The remaining contribution (\ref{wave}),
which has the form of a temperature-dependent wave function
renormalization, is not gauge invariant, and it also suffers from an
infrared divergence. This contribution by itself has no physical
meaning. It combines with previous contributions to form a
gauge-invariant result with no infrared divergence. 

 In parallel with the partition (\ref{decomp}), we write
\begin{equation}
\Delta E^{(T)}({\bf p}) = \Delta E^{(T)}_\gamma({\bf p}) +
	\Delta E^{(T)}_e({\bf p}) \,,
\label{summm}
\end{equation}
with the total energy shift given by Eq.~(\ref{eshift}).  Using
Eq.~(\ref{gammm}) for the photon part gives, after some calculation, 
\begin{equation}
\Delta E^{(T)}_\gamma({\bf p})  = 
{ e^2 \over E } \int {(d^3{\bf k}) \over
(2\pi)^3} \, { 1 \over k} \, f(k) \,,
\label{gamshift}
\end{equation}
where the positive/negative energy has the mass shell values 
$E = \pm E({\bf p})$. 
Similarly, some work with the electron part (\ref{eee}) yields
\begin{eqnarray}
\Delta E^{(T)}_e({\bf p})  &=& 
{ e^2 \over E } \int {(d^3{\bf q}) \over
(2\pi)^3} \, { 1 \over E({\bf q}) } \, n(E({\bf q}))
\Bigg\{ 1
\nonumber\\
&& \qquad\quad + {  m^2 \over \left[ E({\bf q}) + E({\bf p}) \right]^2
- k^2 }  + {  m^2 \over \left[ E({\bf q}) - E({\bf p}) \right]^2
- k^2 }  \Bigg\} \,.
\label{totalll}
\end{eqnarray}
Averaging over the solid angle of ${\bf q}$ produces  
\begin{equation}
\Delta E^{(T)}_e({\bf p}) = { e^2 \over E } \int {(d^3{\bf q}) \over
(2\pi)^3} \, { 1 \over E({\bf q}) } \, n(E({\bf q})) \, 
\Bigg\{ 1 - { m^2 \over 4 \, q  p }
\ln \left( { q + p \over q - p } \right)^2 \bigg\} \,. 
\label{elshift}
\end{equation}
These energy shift corrections are perfectly finite: They have neither
ultraviolet nor infrared divergences.  They also vanish in the zero
temperature limit. Moreover, as remarked before, the energy shifts are
gauge invariant. This gauge invariance may be confirmed by  
explicitly computing the self-energy function in the Feynman
gauge, $\Sigma^{(T)}_F({\bf p},E)$.  This function is obtained by 
making the replacement 
$
\delta_{lm} {-} \hat k_l \hat k_m \to g_{\mu\nu}
$ 
in Eq.~(\ref{f12}), where $g_{\mu\nu}$ is the Minkowski
four-dimensional space-time metric with the spatial gamma matrices
$\gamma_l$, $\gamma_m$ replaced by the space-time matrices
$\gamma_\mu$, $\gamma_\nu$. The Coulomb contribution (\ref{f13}) is now,
of course, omitted. It is a simple matter to evaluate this
replacement. Passing to the mass shell, $ - p^\mu p_\mu = m^2 $, with
effectively $\gamma^\mu \to p^\mu / m $, it is easy to show that
\begin{equation}
\Sigma^{(T)}_F({\bf p},E) \to {E \over m} \, \Delta E^{(T)}({\bf p}) 
\,,
\end{equation}
where 
$ \Delta E^{(T)}({\bf p}) $ is given by Eq's.~(\ref{summm}) -- 
(\ref{totalll}). (Our previous work shows that $E/m$ is the correct
factor to relate the energy shift to the mass-shell self energy.)
Gauge invariance is the reason behind the
cancelation of the $ 1 / {\bf k}^2$ terms when the electron
contribution (\ref{eee}) to the self energy is evaluated on the mass
shell. The resulting energy shifts are the shifts in the positions of
the poles of the thermal electron Green's functions --- they give the
change in the energy needed to create an electron or positron of
momentum ${\bf p}$ in the plasma relative to the energy needed to create
them in the vacuum. 

The energy shifts are physical quantities.
Thus we let them stand as separate corrections and
turn to the temperature-dependent ``wave function renormalization''
contribution (\ref{wave}) which is gauge-dependent and which is
not infrared finite, a contribution that must be combined with the
previous result to obtain a physically relevant contribution to the
neutron rate. 
Some calculation with the photon part (\ref{gammm}) of the
self energy yields
\begin{eqnarray}
{\partial \over \partial E } \left[ E^{-1} \, 
\bar\Sigma^{(T)}_\gamma({\bf p}, E) \right] 
	\Bigg|_{E = \pm E({\bf p})} &=&
\mp {2 \,  e^2 \over \pi^2 } \, {1 \over p} \,
	 \int_0^\infty {dk \over k} \, f(k) \, A \,,
\label{gwave}
\end{eqnarray}
where $A$ is the function defined before in Eq.~(\ref{aaa}) but with
different labels,
\begin{equation}
A = 
\left[ 2 \,  E(p)^2 + k^2 \right]
	\ln\left({  E(p) + p \over E(p) - p } \right)
		- 4 \, p \,  E(p)  \,,
\end{equation}
The electron part (\ref{eee}) of the self energy yields
\begin{eqnarray}
&& {\partial \over \partial E } \left[ E^{-1} \, 
 \bar\Sigma^{(T)}_e({\bf p}, E) \right] 
	\Bigg|_{E = \pm E({\bf p})} = \pm
{2 \,  e^2 \over \pi^2 } \,  \int_0^\infty dq \,{q^2 \over E(q)} \, 
n(E(q)) \, { E(p) \over E^2(q) - E^2(p) }
\nonumber\\
&& \qquad\qquad\qquad\qquad
\Bigg\{ 2 + { E^2(q) \over pq } \, \ln\left( { p+q \over p-q }
\right)^2 
\nonumber\\
&& \qquad\qquad\qquad
- { E(q) \over E(p) }  \, { E^2(q) + E^2(p)
\over 2pq} \, \ln\left( { \left[ E(p) E(q) + pq \right]^2 - m^4 \over
	\left[ E(p) E(q) - pq \right]^2 - m^4 } \right) 
\Bigg\} \,.
\label{eslefe}
\end{eqnarray}

For the photon contribution (\ref{gwave}) to the wave function term
(\ref{wave}), we integrate over $E$ and make the simple notational
change ${\bf p} \to {\bf q}$ with, as before, 
${\cal E} = E({\bf q})$. Then, since $qdq = {\cal E} d {\cal E}$ and,
integrating over solid angle,
\begin{equation}
\int { (d^3{\bf p}) \over (2\pi)^3 } \, 
{ \pi \over 2 E({\bf p}) }
\, \cdots	
= { 1 \over 4\pi } \int_m^\infty q \, d {\cal E} 
\, \cdots \,.
\end{equation}
Thus, the photon contribution to the wave function correction
(\ref{wave}) is 
\begin{eqnarray}
\Gamma_{n \, {\rm e-e}}^{\gamma,Z} &=& 
2\, {e^2  G_W^2 \over (2\pi)^5 } \left( g_V^2 + 3 g_A^2 \right)
 \int_0^\infty {dk \over k} \, f(k) \, \int_m^\infty d{\cal E} 
\, A \, \left\{ - 2  \chi({\cal E}) 
	- 2  \chi(- {\cal E}) \right\} 
\nonumber\\
&=&
2\, {e^2  G_W^2 \over (2\pi)^5 } \left( g_V^2 + 3 g_A^2 \right)
 \int_0^\infty {dk \over k} \, f(k) \, \int_m^\infty d{\cal E} 
\, A \, \left\{ - 2  n(-{\cal E}) \, \tilde\chi({\cal E}) 
	- 2  n({\cal E}) \, \tilde\chi(- {\cal E}) \right\} \,.
\label{photon}
\end{eqnarray}
As discussed in the text, this adds simply with the real photon 
contribution Eq.~(\ref{yetinfr}) to produce an infrared finite result.
The second wave function contribution comes from placing 
Eq.~(\ref{eslefe}) in 
Eq.~(\ref{wave}). Removing the first energy integral by the delta
functions, performing the angular integral, and changing integration
variables from momentum to energy with the notational change $q \to
p'$, $E(q) \to E'$, gives the result
\begin{eqnarray}
\Gamma_{n \, {\rm e-e}}^{n} &=&
2 \, { e^2 G_W^2 \over (2\pi)^5 } \, \left( g_V^2 + 3 g_A^2 \right)
\int_m^\infty dE \, \int_m^\infty dE' \,  n(E') \, 
\Big[ \chi(E) + \chi(-E) \Big] 
\nonumber\\
&& \qquad 
 { E \over {E'}^2 - E^2 }
\Bigg\{ 4\,  p p'  +  2 \, {E'}^2  \, \ln\left( { p+p' \over p-p' }
\right)^2 
\nonumber\\
&& \qquad\qquad
-  {E' \over E }  \,  \left [ {E'}^2 + E^2 \right]
 \, \ln\left( { \left[ E E' + pp' \right]^2 - m^4 \over
	\left[ E E' - pp' \right]^2 - m^4 } \right) 
\Bigg\} \,.
\label{waveg}
\end{eqnarray}
This piece by itself is not gauge invariant. As discussed in the text,
adding it to 
Coulomb interaction between the electron and proton, the rate
contribution (\ref{elpro}), 
produces a well-defined, gauge-invariant contribution.

\end{document}